%% file: main.tex
\documentclass[sigconf]{acmart}

\include{utils}

\AtBeginDocument{%
  }

\copyrightyear{2026}
\acmYear{2026}
\setcopyright{cc}
\setcctype{by}
\acmConference[CHI '26]{Proceedings of the 2026 CHI Conference on Human Factors in Computing Systems}{April 13--17, 2026}{Barcelona, Spain}
\acmBooktitle{Proceedings of the 2026 CHI Conference on Human Factors in Computing Systems (CHI '26), April 13--17, 2026, Barcelona, Spain}
\acmPrice{}
\acmDOI{10.1145/3772318.3791794}
\acmISBN{979-8-4007-2278-3/2026/04}

\begin{document}

\title{Auditorily Embodied Conversational Agents: \\ Effects of Spatialization and Situated Audio Cues on Presence and Social Perception}

\author{Yi Fei Cheng}
\authornote{Both authors contributed equally to this research.}
\orcid{0000-0002-6027-4236}
\affiliation{
\department{Human-Computer Interaction Institute}
  \institution{Carnegie Mellon University}
  \city{Pittsburgh}
  \state{Pennsylvania}
  \country{USA}
}
\email{yifeic2@andrew.cmu.edu}

\author{Jarod Bloch}
\authornotemark[1]
\orcid{0009-0009-0012-8649}
\affiliation{
  \institution{Carnegie Mellon University}
  \city{Pittsburgh}
  \state{Pennsylvania}
  \country{USA}
}
\email{jarodb@andrew.cmu.edu}

\author{Alexander Wang}
\orcid{0009-0001-4353-4737}
\affiliation{
\department{Human-Computer Interaction Institute}
  \institution{Carnegie Mellon University}
  \city{Pittsburgh}
  \state{Pennsylvania}
  \country{USA}
}
\email{aw4@andrew.cmu.edu}

\author{Andrea Bianchi}
\orcid{0000-0002-7500-7974}
\affiliation{
  \department{Industrial Design}
  \institution{KAIST}
  \city{Daejeon}
  \country{Republic of Korea}
}
\email{andrea@kaist.ac.kr}

\author{Anusha Withana}
\orcid{0000-0001-6587-1278}
\affiliation{
\department{School of Computer Science}
  \institution{University of Sydney}
  \city{Sydney}
  \country{Australia}
}
\email{anusha.withana@sydney.edu.au}

\author{Anhong Guo}
\orcid{0000-0002-4447-7818}
\affiliation{
    \department{Computer Science and Engineering}
  \institution{University of Michigan}
  \city{Ann Arbor}
  \state{Michigan}
  \country{USA}
}
\email{anhong@umich.edu}

\author{Laurie M. Heller}
\orcid{0000-0002-4735-5701}
\affiliation{
    \department{Department of Psychology}
  \institution{Carnegie Mellon University}
  \city{Pittsburgh}
  \state{Pennsylvania}
  \country{USA}
}
\email{laurieheller@cmu.edu}

\author{David Lindlbauer}
\orcid{0000-0002-0809-9696}
\affiliation{
    \department{Human-Computer Interaction Institute}
  \institution{Carnegie Mellon University}
  \city{Pittsburgh}
  \state{Pennsylvania}
  \country{USA}
}
\email{davidlindlbauer@cmu.edu}

\renewcommand{\shortauthors}{Cheng~\etal}
\renewcommand{\shorttitle}{Auditorily Embodied Conversational Agents}

\begin{teaserfigure}
  \centering
  \includegraphics[width=\linewidth]{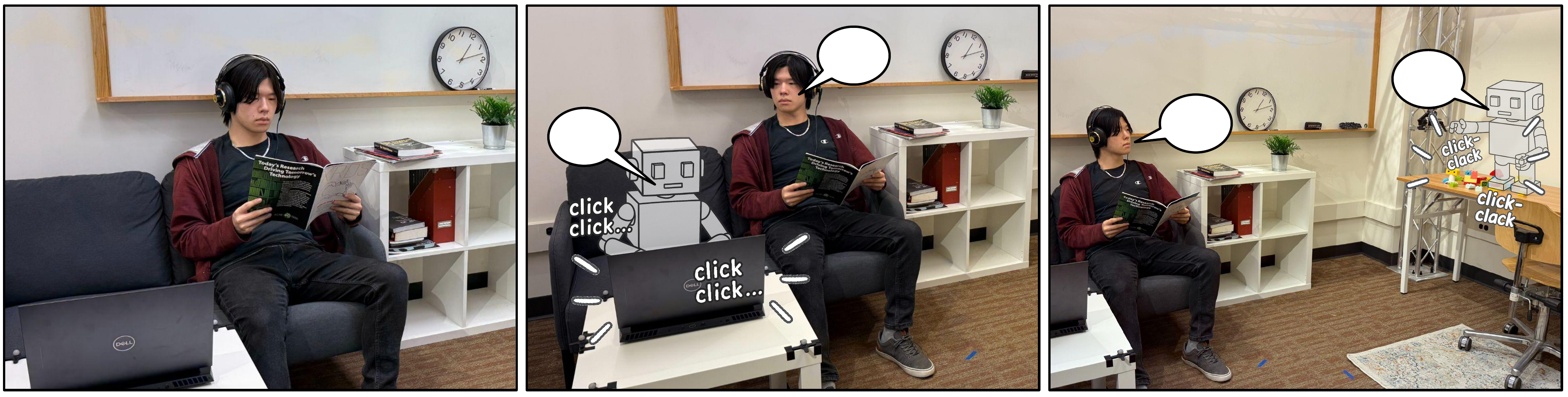}
  \caption{
  We explore how \emph{auditory embodiment} influences perceptions of conversational agents. 
  While agents are often embodied visually, such modalities may not always be available, for instance, when interacting through headphones \textsl{(left)}. 
  We investigate whether embodiment can be conveyed solely through audio, using spatialized voice and situated Foley sounds. 
  For example, an agent may be represented as seated next to the user while typing on a laptop \textsl{(middle)}, or as picking up toy blocks across the room \textsl{(right). The visual depiction of the agent is for illustration purposes only.}
  }
  \Description{A three-panel illustration showing a participant interacting with an auditorily embodied conversational agent in their living room. In the left panel, the participant sits on a couch, reading a magazine. In the middle panel, the participant continues reading while casually conversing with the agent, whose voice is spatialized to sound as if it is sitting next to them; Foley audio of typing reinforces its presence. In the right panel, the agent’s voice and Foley cues shift across the room as it interacts with toy blocks on a table, conveying its movement and activities even though it is represented only through sound.}
\end{teaserfigure}

\input{sections/abstract}

\ccsdesc[500]{Human-centered computing~Empirical studies in HCI}
\ccsdesc[500]{Human-centered computing~Sound-based input / output}

\keywords{Presence, Embodiment,  Agents, Spatial Audio}

\maketitle

\input{sections/introduction}
\input{sections/background}
\input{sections/implementation}
\input{sections/experiment}
\input{sections/results}

\input{sections/qualitative}
\input{sections/discussion}
\input{sections/conclusion}
\input{sections/acknowledgements}

\bibliographystyle{ACM-Reference-Format}
\bibliography{references}

\input{sections/appendix}

\end{document}

%% file: utils.tex
\usepackage{xspace}
\usepackage{soul}
\newcommand{\etal}{et~al.\xspace}
\newcommand{\ie}{i.e.,\xspace}
\newcommand{\etc}{etc.}
\newcommand{\eg}{e.g.,\xspace}

\newcommand{\statsum}[2]{{\small $M=#1$, $SD=#2$}}
\newcommand{\sd}[1]{{\small $SD=#1$}}
\newcommand{\pval}[2]{{\small $p\,#1\,#2$}}
\newcommand{\cronbach}[1]{{\small $\alpha=#1$}}
\newcommand{\art}[6]{{\small $F_{#1,#2}$\,$=$\,$#3$, $p$\,$#4$\,$#5$, $\eta_{p}^{2}$\,$=$\,$#6$}} 

\newcommand\rrh[1]{\textcolor{black}{#1}}

%% file: sections/abstract.tex
\begin{abstract}
Embodiment can enhance conversational agents, such as increasing their perceived presence. 
This is typically achieved through visual representations of a virtual body;
however, visual modalities are not always available, such as when users interact with agents using headphones or display-less glasses. 
In this work, we explore \emph{auditory embodiment}. 
By introducing auditory cues of bodily presence -- through spatially localized voice and situated Foley audio from environmental interactions -- we investigate how audio alone can convey embodiment and influence perceptions of a conversational agent.
We conducted a 2~(\textsc{spatialization}: \textsl{monaural} vs. \textsl{spatialized}) × 2~(\textsc{Foley}: \textsl{none} vs. \textsl{Foley}) within-subjects study, where participants (n=24) engaged in conversations with agents. Our results show that spatialization and Foley increase co-presence, but reduce users’ perceptions of the agent’s attention and other social attributes.
\end{abstract}

%% file: sections/introduction.tex
\section{Introduction}
In recent years, conversational agents 
(\eg~Google’s Gemini Live, OpenAI’s ChatGPT voice mode)
have advanced rapidly in their ability to engage in naturalistic dialogue.
They increasingly demonstrate human-like behaviors~\cite{cheng2025surrogates} and can respond to spoken inputs in real time~\cite{defossez2024moshi}.
These capabilities have driven the growing popularity of applications and experiences that support conversational interactions, ranging from interactive museum guides~\cite{lopezgarcia2024speakingobjects} to services for social engagement and emotional support~\cite{fang2025aihumanbehaviorsshape}. 

In the design of conversational agents, an important factor shaping user experiences is the way the agent is represented~\cite{nowak2018avatars}. 
A common approach to representation is \emph{embodiment}, where agents are given a bodily form~\cite{cassell2000eca}.
Prior work has shown that embodying an agent can yield a range of advantages, from providing richer multimodal communication cues to support user tasks~\cite{huang2021adaptutar} to enhancing the agent’s perceived social presence~\cite{bainbridge2008presence}.
More importantly, embodiment shapes how people behave around agents, for better or worse.
For instance, embodiment can foster greater trust in the agent~\cite{kim2018body}.
This can be beneficial in domains such as education where trust may support student engagement~\cite{pataranutaporn2021ai}; however,
this can also lead to unrealistic expectations of the agent’s abilities~\cite{de2016ethical}.

These implications have motivated extensive research on how best to represent conversational agents. 
In both prior research and commercial systems, embodiment has typically been realized through visual representations~\cite{norouzi2018iva}, such as presenting avatars in Virtual or Augmented Reality (VR/AR). 
Yet, visual output modalities are not always available or desirable. 
A user may interact with an agent through earphones while their device is in their pocket, such as when walking or cooking. 
Recent smart glasses, such as the Ray-Ban Meta Glasses~\cite{metaaiglasses}, are also display-less and rely primarily on auditory interactions. 
In these scenarios, current methods and understandings of agent embodiment may not be applicable.

In this work, we explore \emph{auditory embodiment} of conversational agents.
Specifically, we ask: \textbf{can embodiment be introduced through audio alone, and how does this influence user perceptions of the agent?}

In real-world social contexts, people are often co-present with others they cannot see, such as when someone is behind them.
Nevertheless, their presence can still be sensed through (1) the spatial location of their voice and (2) incidental sounds generated by bodily movements and interactions with the environment (\eg~footsteps or rustling clothing).
Similarly, in film, auditory realism is often enhanced through the use of everyday \emph{Foley} sounds that accompany on-screen actions.
Drawing on how humans naturally use incidental bodily and environmental sounds to infer others’ presence in shared space, as well as the use of everyday Foley sounds to convey situated actions in film, we investigate whether such auditory cues can serve as mechanisms for supporting the embodiment of conversational agents.
In particular, we examine how \emph{spatialization} and situated \emph{Foley} audio that represent an agent’s movement and actions shape user perceptions and behaviors.

As an initial exploration of \emph{auditory embodiment}, we study its effects in the context of \emph{casual social interactions} with conversational agents.
Prior research has long examined embodied agents in social settings~\cite{cassell2000eca}, with more recent work focusing on conversational agents as companions~\cite{fang2025aihumanbehaviorsshape}.
Our work aims to provide insights into the design of conversational agents for social use, particularly in how they may be represented auditorily.

To this end, 
we conducted a within-subjects controlled study (n=24) in which participants engaged in and evaluated conversations with a conversational agent.
In each conversation, the agent’s voice was rendered either spatialized or monaural (non-spatial).
For each spatialization condition, we further varied whether the agent’s audio was presented alone or accompanied by Foley sounds.
Our results show that \textsl{spatializing} the agent’s audio and adding \textsl{Foley} contribute to stronger feelings of co-presence.
However, the addition of \textsl{Foley} also reduces attention and message comprehension, and leads to a negative social impression.

Overall, we contribute empirical results showing how spatialization and Foley influence the social presence and perception of conversational agents.
Through our results, 
we discuss opportunities and challenges of auditorily embodying conversational agents, highlighting considerations for the design of future systems.

%% file: sections/background.tex
\section{Related Work}
Our study builds on prior research on auditory interfaces, embodied conversational agents, and social presence.

\subsection{Auditory Interfaces}

Over the last 40 years, there has been persistent interest in auditory interfaces and interactions~\cite{gaver1987auditory}. 
Early work on sound-based interactions, such as \emph{earcons}~\cite{blattner1989earcons} and \emph{auditory icons}~\cite{gaver1987auditory}, primarily focused on information delivery. 
Subsequent research has significantly broadened this scope, 
exploring sound as a medium for embodied interaction~\cite{muller2014boomroom}, 
as a customizable component of domestic environments~\cite{jacobsen2023living}, 
and as a means of increasing the accessibility of emerging technologies such as Mixed and Virtual Reality~\cite{jain2021taxonomy,chang2024soundshift}.

Within this broader landscape, the work most closely related to ours concerns \emph{voice interfaces}\footnote{Also referred to as dialogue systems, voice assistants, conversational agents,~\etc}, which use natural language as input and output.
Voice interfaces have been explored across diverse applications, from in-vehicle assistance~\cite{wang2022vadriving} to managing chronic and mental health conditions~\cite{berube2021health}.
According to \citeauthor{rzepka2022voice}~\cite{rzepka2022voice}, 
voice interfaces can provide pragmatic benefits such as convenience, ease of use, and time savings, as well as hedonic and social value.

The user experience of voice interfaces is highly sensitive to various design parameters. 
For example, \citeauthor{clark2019goodconversation}~\cite{clark2019goodconversation} showed that users’ perceptions are shaped by the role of the conversation (\ie~social or functional).
The characteristics of an interface’s voice can also shape the dynamics of an interaction~\cite{cambre2019onevoice} and how it is perceived in terms of gender, age, and personality~\cite{ko2006voice,mcaleer2014voice}.
These prior works informed what variables we needed to control in our own study. 
Drawing on \citeauthor{clark2019goodconversation}~\cite{clark2019goodconversation}, 
we constrained our conversation task to a social role.
Similarly, we fixed both the agent’s voice and its system instructions to encourage more consistent responses.

One design parameter of particular relevance to our work is the spatial positioning of the interface's voice.
In voice calls, prior work has shown that spatializing participants' voices can improve memory, speaker identification, and social presence~\cite{kilgore2003spatialized,dicke2010talk}.
\citeauthor{takyama2010throwingvoices}~\cite{takyama2010throwingvoices} showed that manipulating the spatial position of a conversational agent’s voice can make disagreements seem more palatable. 
\citeauthor{hyrkas2023spatializedaudio}\cite{hyrkas2023spatializedaudio} and \citeauthor{nowak2023hear}~\cite{nowak2023hear} extended these findings to video conferencing, showing that spatial audio increased perceptions of interactivity while reducing cognitive effort.
In augmented and virtual reality, 
prior work has also shown that users generally prefer richer spatialized auditory and visual user representations~\cite{fink2024avatarsvisual,immohr2023multimodalspatialaudio}.
We extend this line of research by asking whether spatializing a conversational agent’s voice can serve as a mechanism for \emph{auditory embodiment}. 
While \citeauthor{kilgore2003spatialized}~\cite{kilgore2003spatialized}, \citeauthor{dicke2010talk}~\cite{dicke2010talk}, and \citeauthor{takyama2010throwingvoices}~\cite{takyama2010throwingvoices} focused on the effects of spatialized audio for virtual sounds that are not anchored in the listener’s physical environment, our work examines whether spatialization can situate an \emph{agent} within the user’s surroundings. 
In contrast to work exploring how spatialized audio complements visual modalities~\cite{hyrkas2023spatializedaudio,nowak2023hear,fink2024avatarsvisual,immohr2023multimodalspatialaudio}, we study whether spatial positioning can convey a sense of bodily presence in audio-only settings.

Another relevant factor is whether the interface produces additional non-speech audio cues.
Prior work shows that emotion-evoking sounds shape perceptions of social attractiveness and emotional states~\cite{hanser2015effects,may1980effects,logeswaran2009music,tao2025audiopersonas}.
In Human-Robot Interaction, ``consequential'' sounds (\ie~by-products of robot mechanics or movements, rather than intentionally designed) can likewise influence evaluations of competence, trustworthiness, and human-likeness~\cite{tennent2017goodvibrations,moore2017intentional}.
In our work, we examine how situated Foley audio shapes user perceptions and behavior. 
Drawing on film, where Foley enhances scene realism, and on how everyday sounds inform spatial and semantic understanding~\cite{gaver1993world}, we ask whether analogous cues can contribute to auditory embodiment for an audio-only agent.
Building on prior findings that sound shapes social perception, we evaluate how such Foley cues affect users’ conversational experience, including social presence, attraction, and likeability.

\subsection{Embodied Conversational Agents}
Embodied conversational agents are dialogue systems that are represented by either a virtual or physical body~\cite{cassell2000eca}.
Substantial literature has examined their design and effects on user experience~\cite{wang2021nonverbal,norouzi2018iva,norouzi2020embodied,nowak2018avatars,yang2025embodiedagents,kyrlitsias2022social}. 
Introducing embodiment can yield a range of advantages.
Functionally, a bodily representation allows agents to convey multimodal cues when communicating with users, such as gestures, gaze, and proxemics~\cite{andrist2017coordinated,huang2022proxemicsagents,zhang2025embodiedagent}. 
Embodiment further affects an agent’s social presence~\cite{kim2018body} (Section~\ref{sec:related-work-presence}), with important consequences for trust, engagement, and related factors~\cite{bainbridge2008presence,hoffman2015robotpresence}. 
Beyond these benefits,
embodiment also changes how people perceive and interact with these systems~\cite{reeves1996media}. 
In line with the \emph{computers as social actors} paradigm~\cite{nass1994socialactors}, when agents are embodied, users tend to treat them more like other people, thereby enabling them to adopt a more social role~\cite{koda1996faces,takeuchi1995facialdisplays}.

A longstanding question in embodied conversational agent research concerns how such agents should be optimally represented~\cite{nass2000researching,cassell2001eca}.
Prior work has typically embodied agents to imitate humans~\cite{hirzle2023xrai}.
However, researchers have also questioned the value of adopting human-like appearances. 
For instance, \citeauthor{hale2024embodiment}~\cite{hale2024embodiment} suggest that physical bodies can inadvertently evoke stereotypes related to gender, ethnicity, and beauty.
Moreover, optimizing for highly realistic representations risks negative evaluations due to the uncanny valley effect~\cite{groom2009realism,kaiser2025getreal}.
Beyond debates over whether agents should appear human, 
prior work has explored a wide range of alternative appearances and behaviors~\cite{dehn2000agent}.
For example, \citeauthor{weber2021foodchattar}~\cite{weber2021foodchattar} investigated user perceptions of agents embodied as food items. 

In our work, we build on this research on examining how conversational agents should be embodied.
In contrast to previous studies that have primarily investigated how different visual embodiments influence user perceptions and task outcomes, we focus on the effects of auditory bodily representations.
It is worth recognizing here that previous work has suggested that the absence of a visual representation is generally detrimental to the user experience~\cite{rasch2025creepycocreator,kim2019patientcare,wang2019va}. 
Yet, visual output modalities may not always be available or appropriate. 
Users may engage with conversations with agents through audio-based wearables~\cite{zulfikar2024memoro} or display-less smart glasses~\cite{metaaiglasses}. 
Moreover, prior work has shown that in some contexts like driving, voice-only agents can also be advantageous for task performance and efficiency~\cite{hajahmadi2025embodied,wang2022vadriving}.  
Therefore, we investigate audio-only manipulations to address these scenarios.

\subsection{Social Presence}
\label{sec:related-work-presence}
Social presence was initially conceptualized by \citeauthor{short1976telecommunication}~\cite{short1976telecommunication} as a characteristic of interpersonal communication, defined as ``the degree of salience of the other person in the interaction and the consequent salience of the interpersonal relationship.''
Numerous studies have shown that social presence is associated with a range of positive communication outcomes, which has motivated a longstanding interest in identifying its antecedents~\cite{oh2018socialpresence}.
Early work examined how social presence is shaped by different communication modalities, such as speakerphone audio, monaural and multichannel audio, video, and face-to-face interaction~\cite{short1976telecommunication}.
Subsequent research explored additional factors, such as the availability of social cues~\cite{kimmel2023faceit}, viewing perspective~\cite{wang2025avatarvisibility}, and user demographics~\cite{kimmel2024kinetic}.

Although originally defined for human interactions, social presence has also been shown to apply to interactions with computational agents. 
Early dialogue systems, such as ELIZA, demonstrated that even a rudimentary text-based interface could elicit responses from users as if they were conversing with a real person~\cite{weizenbaum1966eliza,weizenbaum1976computer,turkle2011alone}. 
As a \emph{social actor}~\cite{nass1994socialactors}, an agent’s perceived social presence can shape how users behave and how they perceive the agent, which has significant implications for emerging application areas such as social interaction and emotional support~\cite{fang2025aihumanbehaviorsshape}.

However, prior work has primarily focused on the effects of visual representations on social presence~\cite{oh2018socialpresence}. 
In contrast, our work examines how audio-only agent representations shape perceived social presence, aiming to understand how to design agent representations that more effectively support better user experiences.

%% file: sections/implementation.tex
\section{Auditory Embodiment: Concept and Implementation}
\label{sec:system}
\emph{Embodiment} refers to introducing a bodily representation that grounds an agent in the user’s environment~\cite{cassell2001eca}.
We extend this notion to \emph{auditory embodiment}, which we conceptualize as the extent to which an agent’s bodily presence is conveyed through sound.

In everyday social contexts, auditory cues contribute to our awareness of others’ bodily presence.
When someone in our environment speaks, our auditory system processes not only the semantic content of their speech but also spatial cues arising from sound propagation, allowing listeners to infer the speaker’s approximate physical location within the shared space~\cite{moore2012introduction}.
This sense of presence is further reinforced by the sounds of their activities with the environment (\eg~walking, placing an object, or striking a surface).
In particular, such \emph{action sounds}~\cite{serafin2018sonic} enable us to situate others relative to an existing cognitive map of the environment~\cite{clemenson2021rethinking}.

Drawing on these observations, our work considers two approaches to achieving \emph{auditory embodiment}:
(1) \emph{spatializing} the agent’s sounds, and
(2) introducing \emph{Foley} sounds that reflect its situated interactions with the environment. 
We implemented these approaches in a Unity-based system that formed the basis of our experimental apparatus. We describe each approach in detail below.

\subsection{Spatialization}
The human auditory system relies on a rich set of perceptual processes for localizing sounds~\cite{moore2012introduction}.
In the horizontal plane, sound localization primarily depends on interaural time differences~(ITD) and interaural level differences~(ILD).
These cues reflect differences in a signal’s arrival time and sound pressure level at each ear, and vary with the position of sound sources relative to the listener’s head.
Aside from ITD and ILD cues, sound localization is also shaped by the \emph{head-related transfer function} (HRTF)~\cite{wenzel1993localization}, which describes how sound waves are filtered by the anatomical features of the listener (\eg~the shape of the head and outer ears) before perception.

To simulate hearing an agent from a specific location in space, a system can reproduce human sound localization cues by modeling sound propagation from the agent’s pose relative to the listener and rendering binaural signals using an HRTF.

\subsubsection{Implementation}
\label{sec:spatialization-implementation}
Our system simulated the experience of speaking with an agent positioned at a specific location in the user’s room 
by tracking the user’s head orientation relative to the agent and applying spatial audio rendering to reproduce the resulting 3D sound dynamics. 
For this purpose, our system integrated ten OptiTrack cameras to track the user’s head position and rotation using a head-mounted five-marker rigid body.
This setup enabled real-time, head-relative spatial audio updates, ensuring the agent’s perceived location remained stable during user movement.
For HRTF spatial audio rendering, we used the spatializer from the Meta XR Audio SDK version 77.0.0~\cite{metaxraudiosdk}, replicating the approach of \citeauthor{tao2025audiopersonas}~\cite{tao2025audiopersonas}.
At the time of the study, the SDK represented the state-of-the-art in spatial audio rendering~\cite{cho2024auptimize,Cheng2024Hearing}.
To represent the agent, we set up a virtual sound source that used a ``human voice'' directivity pattern~\cite{metaaudiosourcedirectivity}, simulating natural voice attenuation (\ie~quieter and more muffled when turned away from the user).
Lastly, our system applied room acoustic effects to virtual sounds using the SDK’s shoebox model~\cite{metaroomacoustics}, with room dimensions and material properties configured to match the experimental space.

\subsection{Foley}
Almost every bodily movement, such as walking, flipping through a magazine, or typing, produces action sounds that convey spatial information~\cite{tajadura2012action}.
In social scenarios, these sounds support awareness ofothers' whereabouts by enabling associations between their actions and the physical objects involved.

To reinforce an agent’s presence within a user’s space,
a system may reproduce sounds that plausibly correspond to the agent’s interactions with the environment.
In film, this practice is known as \emph{Foley}, where everyday sound effects are created to enhance the perception of actions on screen.
In our context, such sounds can serve as complementary cues to the agent’s speech, helping to contextualize its activities within the shared space.
This approach further parallels how Augmented Reality uses visual augmentations to anchor virtual entities within the user’s environment, with Foley cues serving an analogous role in auditorily situating the agent.

\subsubsection{Implementation}
To simulate incidental sounds associated with the agent’s embodied presence, our system replays recorded audio clips from manually configured spatial locations within the room.
Using the same tracking and audio rendering components described in Section~\ref{sec:spatialization-implementation}, these sounds are rendered as though emanating from contextually appropriate positions (\eg~keyboard clicks from the location of a physical laptop).
Our system currently supports predefined \emph{activity sequences}, each consisting of multiple Foley events bound to a pre-recorded agent movement trajectory.
We discuss the design of the activity sequences in Section~\ref{sec:agent-activity}, including the curated stimuli used in our experiment.
Potential automated, context-aware implementations are considered in Section~\ref{sec:discussion}.

%% file: sections/experiment.tex
\section{Experiment}
\label{sec:experiment}
To investigate how auditory embodiment influences perceptions of conversational agents, we conducted a within-subjects study. 
Each participant engaged in four conversations with the agent, varying along two factors: whether the agent’s audio was spatialized and whether it was accompanied by Foley. 
We measured both subjective perceptions, through ratings of social presence and impressions of the agent, and behavioral responses, including conversational dynamics and user movements within the environment.

\subsection{Conversation Task}
In each condition, participants were tasked with engaging in 3-minute conversations with an agent. 
Considering the potential use case of conversational agents as companions~\cite{fang2025aihumanbehaviorsshape}, 
we designed our conversation task to simulate a casual conversation scenario.
Participants randomly selected a conversation topic from a curated set adapted from \citeauthor{fang2025aihumanbehaviorsshape}~\cite{fang2025aihumanbehaviorsshape}.
We specifically selected self-relevant topics that required some sharing of personal experiences. 
To minimize variability, topics with negative valence or high arousal were excluded. 
Example topics included how participants had celebrated a recent holiday or the best show they had watched recently (see~\autoref{appendix:topics} for the full list).

For each conversation, the agent was configured with the same system instructions that specified it was a companion engaging in casual conversation with the user (\autoref{appendix:voice-config}). 
Through pilot testing, we experimented with several instruction sets, including the engaging voice configuration described by \citeauthor{fang2025aihumanbehaviorsshape}~\cite{fang2025aihumanbehaviorsshape}. 
We ultimately adopted our final configuration, as it qualitatively produced the most consistent responses.

\subsection{Agent Behavior}
\label{sec:agent-activity}
Our design of the agent’s embodied behaviors was guided by the following objectives:

\textbf{Perceptual grounding:} The agent’s activities should involve both movement and recognizable interactions with objects, enabling participants to perceive its spatial location and engagement with the environment. 
In particular, prior research has shown that humans localize sound more effectively through relative changes than through static, absolute cues~\cite{recanzone1998comparison}. 
Therefore, we introduced agent movement to support these relative judgments and strengthen perceptions of the agent as being situated within the environment.

\textbf{Ecological plausibility:} The agent should perform everyday activities that could plausibly occur within the experimental room.

\textbf{Ambient:} The agent’s actions should remain in the background, providing subtle cues of presence.

\begin{figure}[t!]
    \centering
    \includegraphics[width=\linewidth]{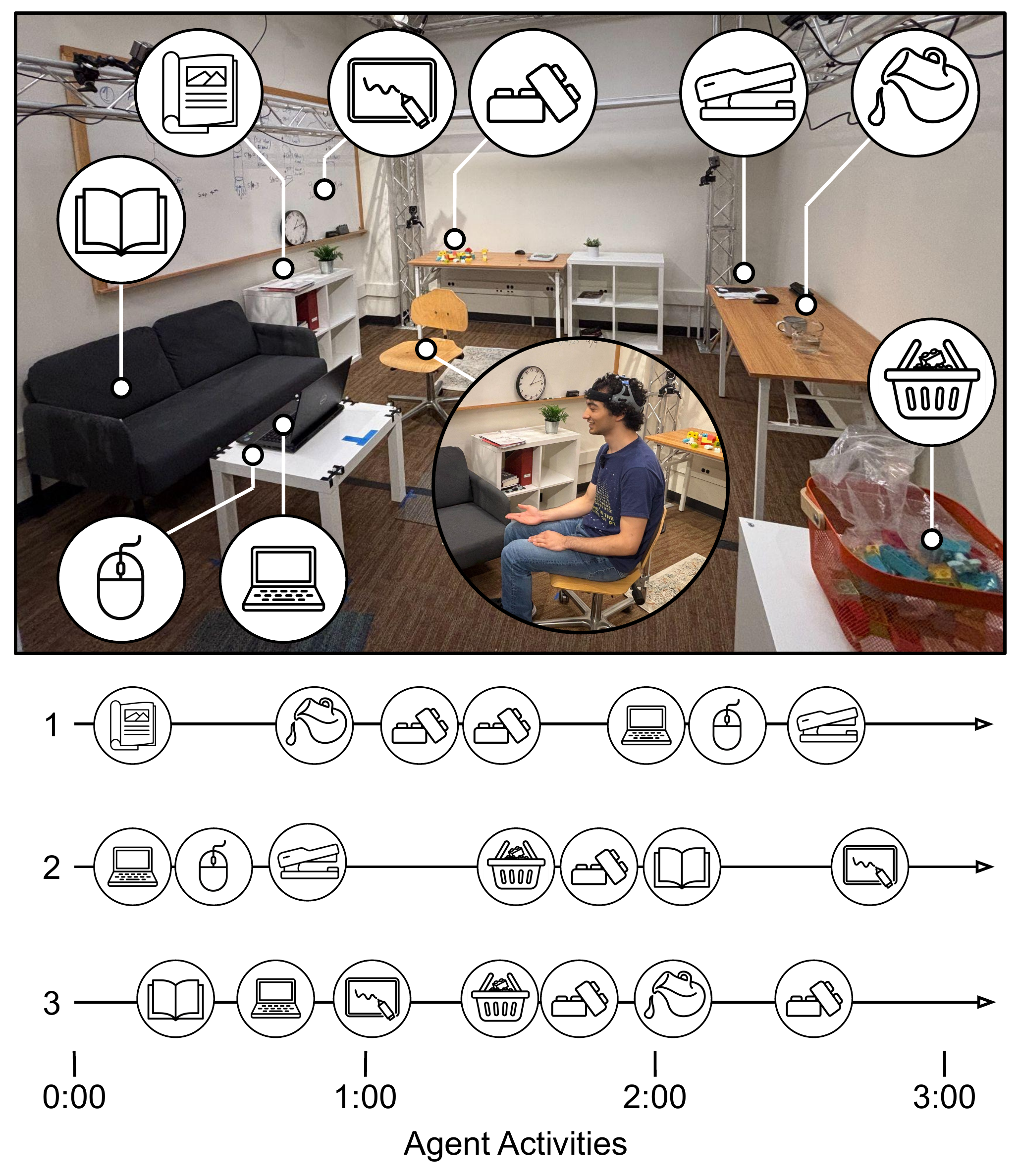}
    \caption{Agent activity sequences. \emph{(Top)} shows the activities and their corresponding locations within the room, with the participant positioned at its center. \emph{(Bottom)} shows the temporal ordering of the three activity sequences the agent followed in our experiment.}
    \Description{
     Figure illustrating agent activities and their temporal sequences. The figure is divided into two panels. The top panel shows a photograph of the experimental room. Nine circular icons are connected by lines to specific locations or objects within the room. A book icon points to a couch, representing reading. Mouse and laptop icons point to their respective devices on a coffee table, representing computer use. A magazine icon points to a magazine on a shelf next to the couch, representing reading. A whiteboard icon points to a whiteboard at the back of the room, representing writing. A blocks icon points to a table in the background, representing block assembly. A stapler icon points to a stapler on the right side of the room, representing stapler use. A pitcher icon points to a cup on the right, representing pouring water. A basket icon points to a basket on the right, representing rummaging through a basket. A circular inset shows a participant seated in the room, indicating the participant’s seating position.
     The bottom panel shows three horizontal timelines labeled 1, 2, and 3, with a time scale ranging from 0:00 to 3:00. Each timeline displays a sequence of icons from the top panel, representing the order and timing of activities. Sequence 1 consists of magazine, pitcher, blocks, blocks, laptop, mouse, and stapler. Sequence 2 consists of laptop, mouse, stapler, basket, blocks, book, and whiteboard. Sequence 3 consists of book, laptop, whiteboard, basket, blocks, pitcher, and blocks.
    }
    \label{fig:action-sequence}
\end{figure}

To model these behaviors, we drew on the persona of a colleague who occasionally moves about and interacts with objects in their room while maintaining a casual conversation.
We designed three \textsc{activity sequences}, each consisting of seven activities (\autoref{fig:action-sequence}).
The activities included flipping through a book or magazine, writing on a whiteboard, assembling toy blocks, pouring a glass of water, organizing and stapling papers, clicking a mouse, and typing.

For each sequence, a member of the research team recorded the trajectory of their head while walking through the room and performing the activities.
This trajectory was then aligned with Foley sounds for the constituent activities, recorded using a Snowball iCE microphone.
These sounds were manually mapped to their contextually relevant locations within the environment. 
In addition, we introduced footstep sounds mapped to the velocity of the agent’s movement, as well as subtle clothing interaction sounds triggered at random intervals, spatially anchored to the feet and waist.
We notably recorded
our own Foley instead of using generative approaches or sourcing from online catalogs, as early experimentation
indicated that these alternatives lacked the quality and acoustic consistency necessary to be convincingly associated with the space.

\subsection{Experimental Design}
\begin{figure}[t]
    \centering
    \includegraphics[width=\linewidth]{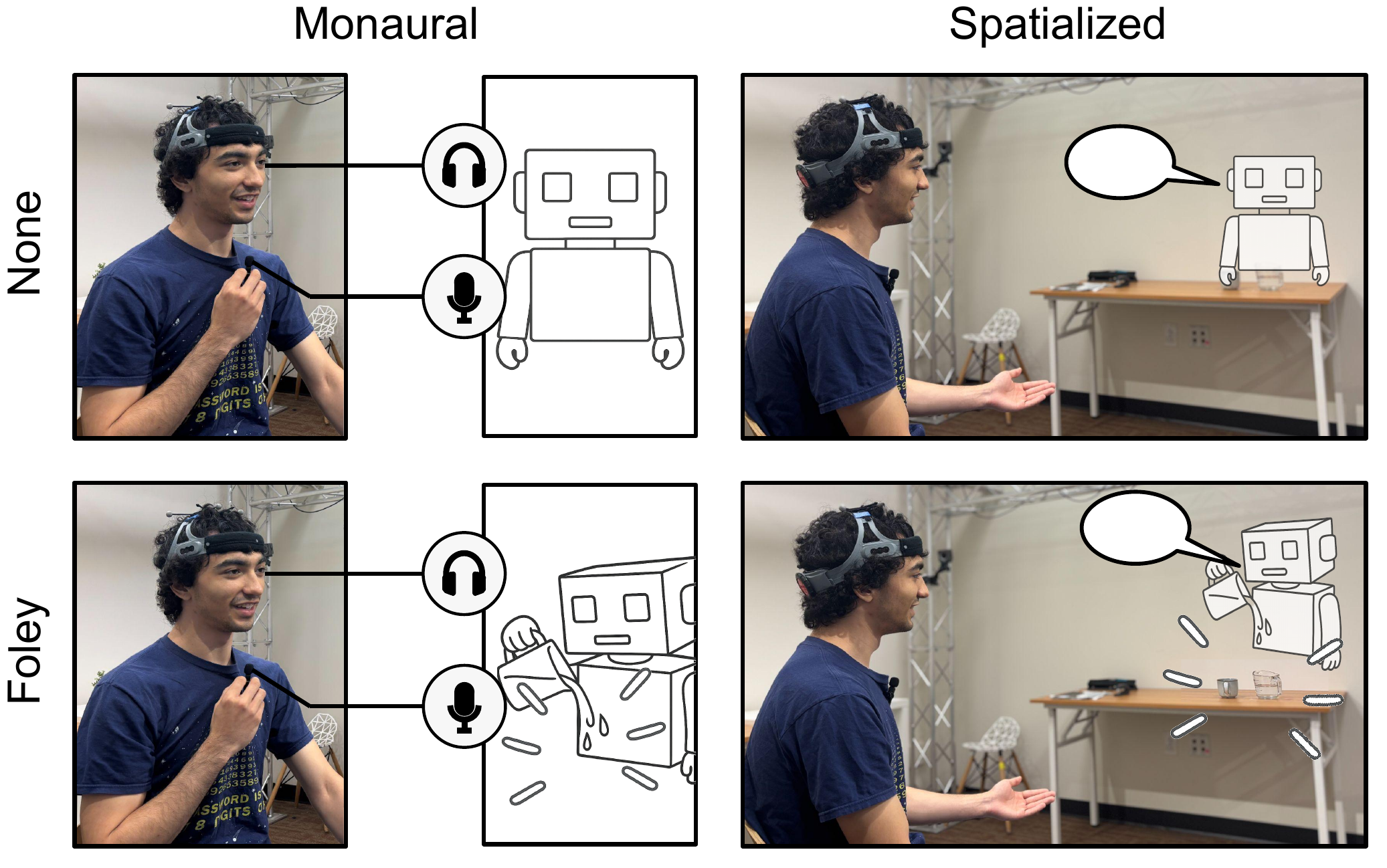}
    \caption{Experimental conditions. The agent’s audio was rendered either \textsl{monaurally} \emph{(left)} or \textsl{spatialized} \emph{(right)}, with the addition of \textsl{Foley} also varied: \textsl{none} \emph{(top)} vs. \textsl{Foley} \emph{(bottom)} (\eg~sounds of the agent pouring water). The visual depiction of the agent is for illustration only; participants experienced all conditions through audio alone.}
    \Description{
    A 2×2 grid illustrating the four experimental conditions of the study. The rows distinguish whether Foley sounds are absent (top) or present (bottom). The columns distinguish whether the agent’s speech is monaural (left) or spatialized (right). Together, the figure depicts: (1) monaural without Foley, (2) spatialized without Foley, (3) monaural with Foley, and (4) spatialized with Foley.
    }
    \label{fig:conditions}
\end{figure}

The experiment followed a $2\times2$ within-subjects design with two independent variables (\autoref{fig:conditions}): \textsc{spatialization} (\textsl{monaural}, \textsl{spatialized}) and \textsc{Foley} (\textsl{none}, \textsl{Foley}).
This yielded four conditions:

\textbf{Monaural + None:} 
In this condition, the agent’s voice is neither spatialized nor accompanied by Foley, effectively serving as a baseline comparable to a standard voice call.

\textbf{Spatialized + None:}
In this condition, the agent’s voice is spatialized but not accompanied by Foley. 
The agent follows the recorded \textsc{activity sequence} trajectory, providing directional cues that indicate its movement and position in space. No additional sounds, such as footsteps or action sounds, are presented.

\textbf{Monaural + Foley:}
In this condition, the agent’s audio is rendered monaurally and accompanied by Foley sounds representing its movements and object interactions. Because neither the speech nor the Foley is spatialized, these cues are not acoustically localized to any position in the room.

\textbf{Spatialized + Foley:}
In this condition, both the agent’s voice and the Foley sounds are spatialized. 
They follow the recorded \textsc{activity sequence} trajectory and the mapped locations of each interaction,
providing directional cues to the agent’s movement as well as to the sounds of its activities within the room.

The order of conditions and the order of the \textsc{activity sequences} were individually counterbalanced using Latin Square designs and then paired.
This resulted in $4~\text{condition orders}~\times~6~\text{sequence orders}$, which we evenly distributed across our 24 participants.

\subsection{Procedure}

Upon arriving at the lab, participants were first given a brief introduction to the study, the equipment involved, and the data we recorded.
Then, they filled out a consent form and a pre-questionnaire. 

Afterwards, participants completed a familiarization phase. 
In this phase, 
they were instructed to walk around the room to get familiar with it. 
In addition, they were instructed to interact with various objects in the room (\eg~writing on the whiteboard, pouring a glass of water), mirroring potential actions of the agent. 
This phase was designed to help participants build a mental model of the room, approximating the level of familiarity they would normally have with their own everyday environments and facilitating associations between the Foley and the corresponding objects.

Participants then proceeded through the conditions of our study. 
In the conversation tasks, participants were instructed to engage with the agent while sitting on a swivel chair that allowed for free head and torso rotation.
They were instructed to remain seated, but could rotate with the chair. 
Because the agent was scripted to move around the environment, allowing participants to walk freely would have introduced substantial variability in the experience. 
We therefore opted for a more stationary setup.
The seat was intentionally placed near the center of the room, next to a coffee table. 
This location simulated a plausible seating position while ensuring that participants could perceive the agent’s activities from all directions.

After each session, participants reported on several subjective metrics in a post-condition questionnaire. 
After all sessions were completed, participants completed a final exit survey and participated in a semi-structured interview, where they discussed  impressions of the agent, and preferred conditions. 
The study was approved by the Institutional Review Board (IRB) of Carnegie Mellon University. 
The full study took 60 minutes. All participants were compensated \$15 for their time. 

\subsection{Apparatus}
\begin{figure}[t]
    \centering
    \includegraphics[width=\linewidth]{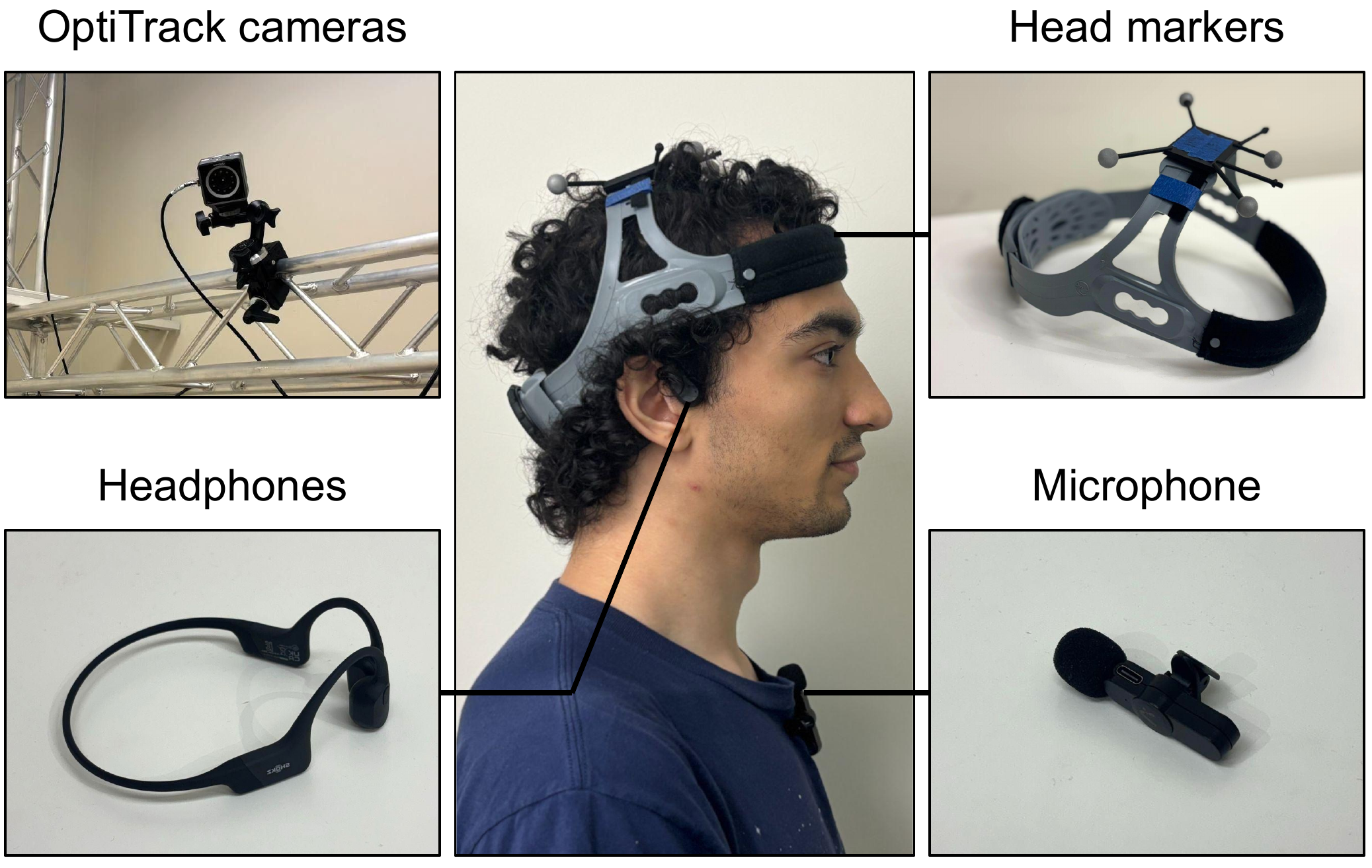}
    \caption{Study apparatus. We used ten OptiTrack motion-capture cameras (\emph{top left}) to track a 4-marker rigid body attached to participants’ heads (\emph{top right}). Participants engaged with the conversational agent using Shokz OpenRun Pro bone-conduction headphones (\emph{bottom left}) and an Aisizon wireless lavalier microphone (\emph{bottom right}).}
    \Description{
    A schematic illustration of the experimental apparatus. The participant is equipped with bone conduction headphones and a wireless lapel microphone. A rigid body with reflective markers is attached to the participant’s head for motion tracking. Around the room, ten OptiTrack cameras are positioned to capture head movement and spatial orientation.
    }
    \label{fig:apparatus}
\end{figure}

The study was conducted in a $4\times3\times3$m experimental space implementing the system described in Section~\ref{sec:system}.
To support interactive dialogue, a bidirectional audio stream was established over WebSocket to a Gemini 2.5 Flash Native Audio model.
The experiment ran on an Intel Core i7-12700H CPU 2.30 GHz computer with 16 GB of RAM, supported by an NVIDIA GeForce RTX 3060 GPU.
Real-time speech input was supported using an Aisizon wireless lavalier microphone and Shokz OpenRun Pro bone-conduction headphones.
We used the Shokz OpenRun headset to preserve environmental sound cues, as prior work suggests that acoustic transparency enhances the real-world grounding of virtual audio~\cite{mcgill2020acoustic}.
Figure~\ref{fig:apparatus} shows an overview of our study apparatus.

\subsection{Measures}
We evaluate participants' perceptions and behaviors through a range of self-reported and quantitative metrics. 
All self-reported metrics were evaluated with 7-point Likert scales (\autoref{appendix:questionnaire-items}). 

\subsubsection{Social Presence}
We assessed participants' perceived social presence of the conversational agent using questions from the Networked Minds Social Presence Inventory (NMSPI)~\cite{biocca2003toward,biocca2003networked,harms2004internal}. 
The NMSPI conceptualizes social presence as comprising multiple sub-dimensions~\cite{harms2004internal}. In our study, we evaluated the following: 

\textbf{Co-presence:} Measures the extent to which users perceive the agent as sharing the same environment. This includes both the user’s own sense of co-presence (perception of self) and their perception of the agent’s co-presence (perception of other).

\textbf{Attentional Allocation:} 
Measures the extent to which users direct attention toward the agent (perception of self) and perceive the agent as allocating attention toward them (perception of other).

\textbf{Message Understanding:}
Measures the user's ability to understand the agent’s messages (perception of self) as well as their perception of the agent’s ability to understand their own messages (perception of other).

\textbf{Affective Understanding:}
Measures the user's ability to recognize the agent’s emotional and attitudinal states (perception of self) and their perception of the agent’s ability to recognize their own emotional and attitudinal states (perception of other).

\textbf{Affective Interdependence:}
Measures the extent to which the user's emotional and attitudinal states influence, and are influenced by, those of the agent. This can be subdivided into perception of self and perception of other.

We excluded the \emph{behavioral interdependence} measure, as our study focused on conversational experience and did not involve joint task performance with the agent.

\subsubsection{Social Impression}
Inspired by prior work~\cite{tao2025audiopersonas,fauville2022impression}, we also assessed participants’ social impressions of the agent through measures of likeability judgments and social attraction:

\textbf{Likeability judgments:} Measured with five items evaluating attractiveness, competence, extroversion, likeability, and trustworthiness, adapted from~\cite{tao2025audiopersonas}.

\textbf{Social attraction:} Measured with four items evaluating friendliness, interpersonal affinity, and willingness to interact again~\cite{fauville2022impression}.

\subsubsection{Preference}
After completing all conversation tasks, participants were asked to rank the four conditions by overall \textsl{preference}. For this ranking, participants were not informed which experimental manipulations each session corresponded to.

\subsubsection{Behavioral measures}
As measures of conversational engagement~\cite{cheng2025surrogates} 
we recorded the \emph{total number of words} exchanged in each conversation (with separate counts for \emph{user words} and \emph{agent words}) and
the number of \emph{turn shifts} (\ie~when the user yielded the floor to the agent and vice versa). 
All verbal metrics were computed from transcriptions generated with OpenAI’s Whisper-Medium~\cite{radford2022whisper}.

As non-verbal behavior indicators of social presence~\cite{kimmel2023faceit},  
we recorded the user's \emph{head rotation} (\ie~the cumulative angular rotation of the participant's head throughout the session) and 
\emph{facing angle to agent} (\ie~the angular difference between the participant's facing direction and the agent’s position relative to them). 
We additionally calculated the percentage of the session in which the agent was positioned within the participant's 
\emph{central} (\ie~$<30^\circ$ of facing direction),
\emph{near-peripheral} (\ie~$<60^\circ$),
and \emph{far-peripheral} vision ($<100^\circ$).
All non-verbal measures were recorded with the OptiTrack system at $\sim20$ Hz.

\subsection{Power and Experimental Participants}
Prior to conducting the study, we performed an a priori power analysis using G*Power 3.1~\cite{faul2009statistical}.
To estimate the required sample size, we considered two effect sizes, $f=0.25$ (small) and $f=0.5$ (medium).
We set the significance level at $\alpha = 0.05$ and the statistical power at $0.8$. Because our subjective measures were collected once per condition, we specified 4 measurements (corresponding to the 4 within-subject conditions) and left the default correlation among repeated measures at 0.5.
The analysis indicated that detecting a small effect would require 24 participants, while detecting a medium effect would require 8 participants. We also considered prior studies on embodiment effects in perceptions of virtual agents (\eg~\cite{wang2019va}).

We recruited 24 participants (11 male, 13 female) between the ages of 18 and 34 (\statsum{27}{5}) from a university community via message groups, social networks, and word-of-mouth.
Most participants (n=23) reported using headphones or earbuds daily, while one reported using them at least once per week.
Participants’ familiarity with relevant technologies is summarized in \autoref{tab:familiarity}.

\begin{table}[t]
\centering
\caption{Self-reported familiarity (number of participants) with voice-based agents, text-based agents, and spatial audio.}
\label{tab:familiarity}
\begin{tabular}{lccc}
\toprule
\textbf{Usage} & \textbf{Voice agents} & \textbf{Text agents} & \textbf{Spatial audio} \\
\midrule
$<$ 5 hours     & 11 & 1 & 2 \\
5--10 hours     & 6  & 3 & 0 \\
10--20 hours    & 3  & 1 & 0 \\
20--30 hours    & 2  & 5 & 4 \\
50--100 hours   & 1  & 5 & 1 \\
$>$ 100 hours   & 1  & 9 & 17 \\
\bottomrule
\end{tabular}
\Description{
Table showing self-reported familiarity (number of participants) with voice-based agents, text-based agents, and spatial audio, organized by cumulative usage hours. For voice-based agents: 11 had <5 hours, 6 had 5–10, 3 had 10–20, 2 had 20–30, 1 had 50–100, and 1 had >100 hours. For text-based agents: 1 had <5 hours, 3 had 5–10, 1 had 10–20, 5 had 20–30, 5 had 50–100, and 9 had >100 hours. For spatial audio: 2 had <5 hours, 0 had 5–10, 0 had 10–20, 4 had 20–30, 1 had 50–100, and 17 had >100 hours.
}
\end{table}

%% file: sections/results.tex
\section{Results}
We evaluated the agent’s social presence, as well as participants’ impressions and preferences, across two levels of \textsc{spatialization} and two levels of \textsc{Foley}. 
In addition, we examined the effects of these factors on participants’ verbal and nonverbal behaviors.
Overall, \textsl{spatializing} the agent’s audio and adding \textsl{Foley} increased participants’ feelings of co-presence.
However, the addition of \textsl{Foley} also reduced attention and message understanding, and contributed to a more negative social impression.

For effect analysis, we analyzed ordinal data (questionnaire ratings) using an Aligned Rank Transform (ART) ANOVA~\cite{wobbrock2011art}. 
Interval data (\eg~words exchanged, turn shifts) were analyzed using a two-factor repeated-measures ANOVA. 
For each dependent variable, \emph{participant} was treated as a random factor, with \textsc{spatialization} and \textsc{audio cues} as within-subject independent variables. 
When assumptions of normality of residuals or homogeneity were violated (Shapiro–Wilk test, $p < .05$), we analyzed the data using ART. 
Post-hoc tests with Bonferroni adjustments
were conducted as needed.
The analysis was performed using R 4.5.1~\cite{rstats}.

\subsection{Social Presence}
\begin{figure}[t]
    \centering
    \includegraphics[width=\linewidth]{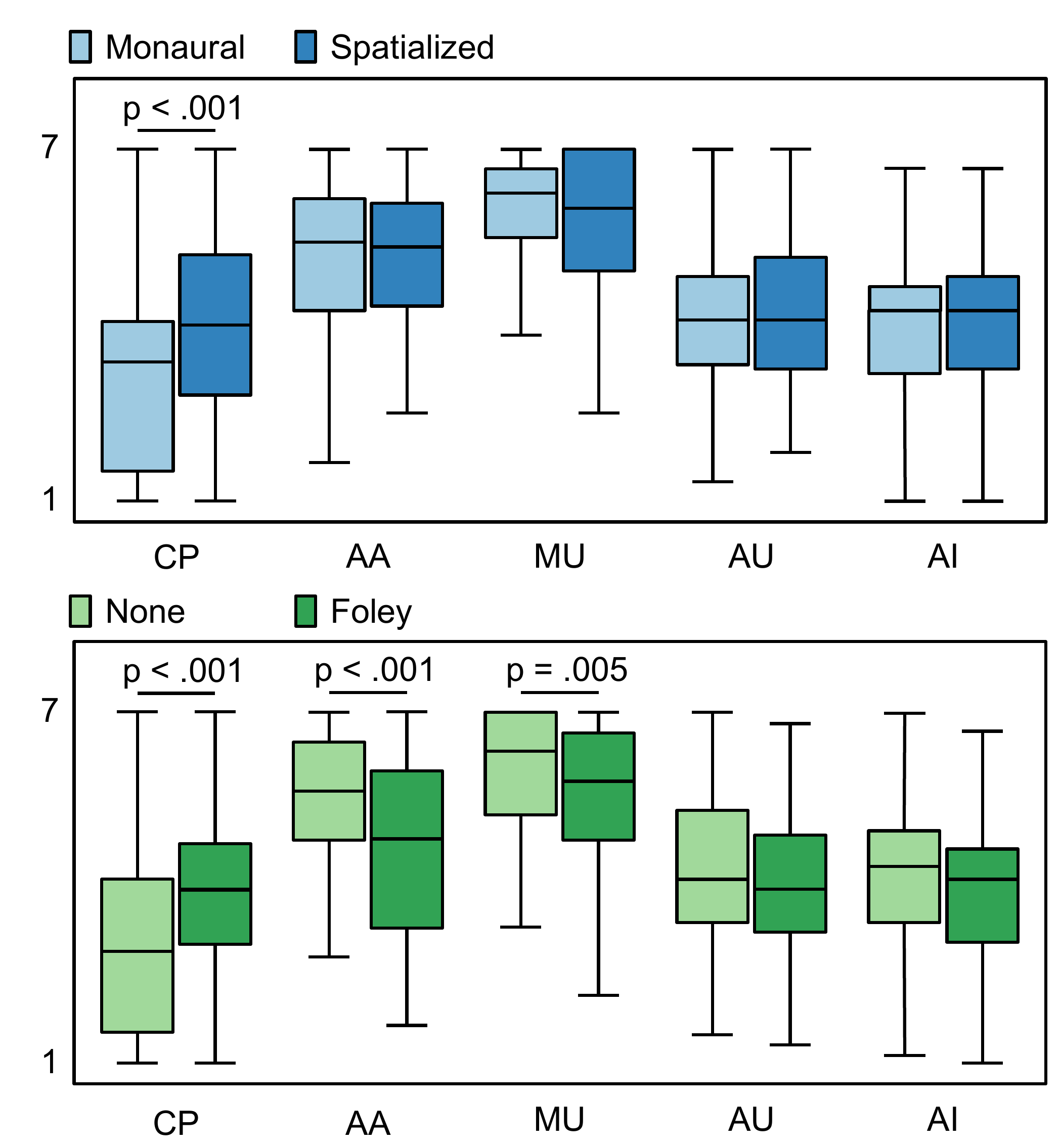}
    \caption{Effect of \textsc{spatialization} (\emph{top}) and \textsc{Foley} (\emph{bottom}) on \emph{social presence}: \emph{co-presence} (CP), \emph{attention allocation} (AA), \emph{message understanding} (MU), \emph{affective understanding} (AU), and \emph{affective interdependence} (AI).}
    \Description{
    Two boxplot panels showing ratings on a scale from 1 to 7 for five social presence categories: Co-presence (CP), Attention Allocation (AA), Message Understanding (MU), Affective Understanding (AU), and Affective Interdependence (AI). 
    The top chart compares Monaural versus Spatialized conditions. The Spatialized condition received significantly higher ratings for Co-presence (CP) than Monaural (p < .001). The other four measures show no significant differences.
    The bottom chart compares Foley conditions with None. Significant differences are marked for three categories. 
    For Co-presence (CP), Foley rated significantly higher than None (p < .001). However, for Attention Allocation (AA) and Message Understanding (MU), the None condition was rated significantly higher than Foley (AA: p < .001; MU: p = .005).
    The remaining measures did not show significant differences. 
    }
    \label{fig:spatial-social-presence}
\end{figure}

The \emph{social presence} factors were each analyzed separately for participants’ \textsl{perception of self}, \textsl{perception of other}, and for a \textsl{combined} score~\cite{biocca2003guide}.
Figure \ref{fig:spatial-social-presence} summarizes the effects of \textsc{spatialization} and \textsc{Foley} on \emph{social presence}.

\subsubsection{Co-presence}
The ART analysis showed a significant main effect of \textsc{spatialization} on \textsl{self} (\art{1}{69}{16.11}{<}{0.001}{0.19}), 
\textsl{other} (\art{1}{69}{5.11}{=}{0.03}{0.07}), 
and \textsl{combined co-presence} (\art{1}{69}{12.66}{<}{0.001}{0.16}). 
Across all three measures, participants reported higher ratings in \textsl{spatialized} conditions compared to \textsl{monaural}, suggesting that \textbf{rendering the agent’s voice spatially can increase feelings of co-presence.}

A main effect of \textsc{Foley} was also observed for \textsl{self} (\art{1}{69}{20.38}{<}{0.001}{0.23}) 
and \textsl{combined co-presence} (\art{1}{69}{12.72}{<}{0.001}{0.16}). 
Participants reported higher ratings in \textsl{Foley} conditions compared to \textsl{none}, suggesting that \textbf{introducing Foley can similarly increase feelings of co-presence.}

No main effect of \textsc{Foley} was found for \textsl{co-presence perception of other} (\art{1}{69}{3.44}{=}{0.07}{0.05}).
No significant interaction effects were found across all \textsl{co-presence} measures (all \pval{>}{0.05}).  

To further investigate the main effects of \textsc{spatialization} and \textsc{Foley}, 
we conducted an exploratory analysis~\cite{johnson2023unmapped,tao2025audiopersonas} probing whether the combined condition (\textsl{spatialized + Foley}) yielded higher ratings than conditions in which only one feature was present (\ie~\textsl{spatialized + none}, \textsl{monaural + Foley}). 
Using Bonferroni-corrected Wilcoxon signed-rank tests, we found that for \textsl{self co-presence}, the \textsl{spatialized + Foley} condition yielded significantly higher ratings than both \textsl{monaural + Foley} (\pval{=}{0.041}) and \textsl{spatialized + none} (\pval{=}{0.004}). \textsl{Combined co-presence} was also significantly higher in \textsl{spatialized + Foley} than in \textsl{monaural + Foley} (\pval{=}{0.02}), though not compared to \textsl{spatialized + none} (\pval{>}{0.05}). These results suggest that \textbf{spatialization and Foley may be complementary in creating a stronger sense of co-presence}.

\subsubsection{Attentional Allocation}
The ART analysis showed a significant main effect of \textsc{Foley} on \textsl{self} (\art{1}{69}{18.40}{<}{0.001}{0.21}), \textsl{other} (\art{1}{69}{8.14}{=}{0.006}{0.11}), and \textsl{combined attentional allocation} (\art{1}{69}{20.10}{<}{0.001}{0.23}). Across all three measures, participants reported lower ratings in \textsl{Foley} compared to \textsl{none}. These results suggest that \textbf{Foley reduced participants’ perceptions of attention between themselves and the agent}.

There was no main effect of \textsc{spatialization} (all \pval{>}{0.05}), 
but it did interact significantly with \textsc{Foley} for both \textsl{self} (\art{1}{69}{4.98}{=}{0.03}{0.07}) 
and \textsl{combined attentional allocation} (\art{1}{69}{7.02}{=}{0.01}{0.09}).
Post-hoc tests indicated \textsl{monaural + Foley} 
reduced \textsl{self} (\pval{=}{0.002}) and \textsl{combined attentional allocation} (\pval{<}{0.001}
compared to \textsl{monaural + none}. 
The \textsl{spatialized + Foley} condition also reduced \textsl{self} (\pval{=}{0.02}) and \textsl{combined attentional allocation} (\pval{=}{0.02})
compared to \textsl{spatialized + none}. 
These findings suggest that \textsc{spatialization} alone did not influence \textsl{attentional allocation}.
The addition of \textsl{Foley} consistently reduced participants’ perceptions of shared \textsl{attention}, regardless of whether the agent’s voice was rendered \textsl{monaurally} or \textsl{spatially}.

\subsubsection{Message Understanding}
We observed a significant main effect of \textsc{Foley} on \textsl{self} (\art{1}{69}{9.40}{=}{0.003}{0.12}),
\textsl{other} (\art{1}{69}{6.81}{=}{0.01}{0.09}),
and \textsl{combined message understanding} (\art{1}{69}{8.28}{=}{0.005}{0.11}).
Participants reported higher ratings across all three measures in conditions 
with \textsl{Foley} compared to \textsl{none}.
These results indicate that \textbf{introducing situated audio cues reduced both participants’ understanding of the agent and their feeling of being understood by the agent.}

No main effects of \textsc{spatialization} or interactions were observed for \textsc{message understanding} (all \pval{>}{0.05}).

\subsubsection{Affective Understanding and Interdependence}
No main effects or interaction effects were observed for \textsc{affective understanding} or \textsc{affective interdependence} (all \pval{>}{0.05}).
These results suggest that within the given experimental setting, spatialization and audio cues did not affect participants’ affective experience.

\begin{figure}[h]
    \centering
    \includegraphics[width=\linewidth]{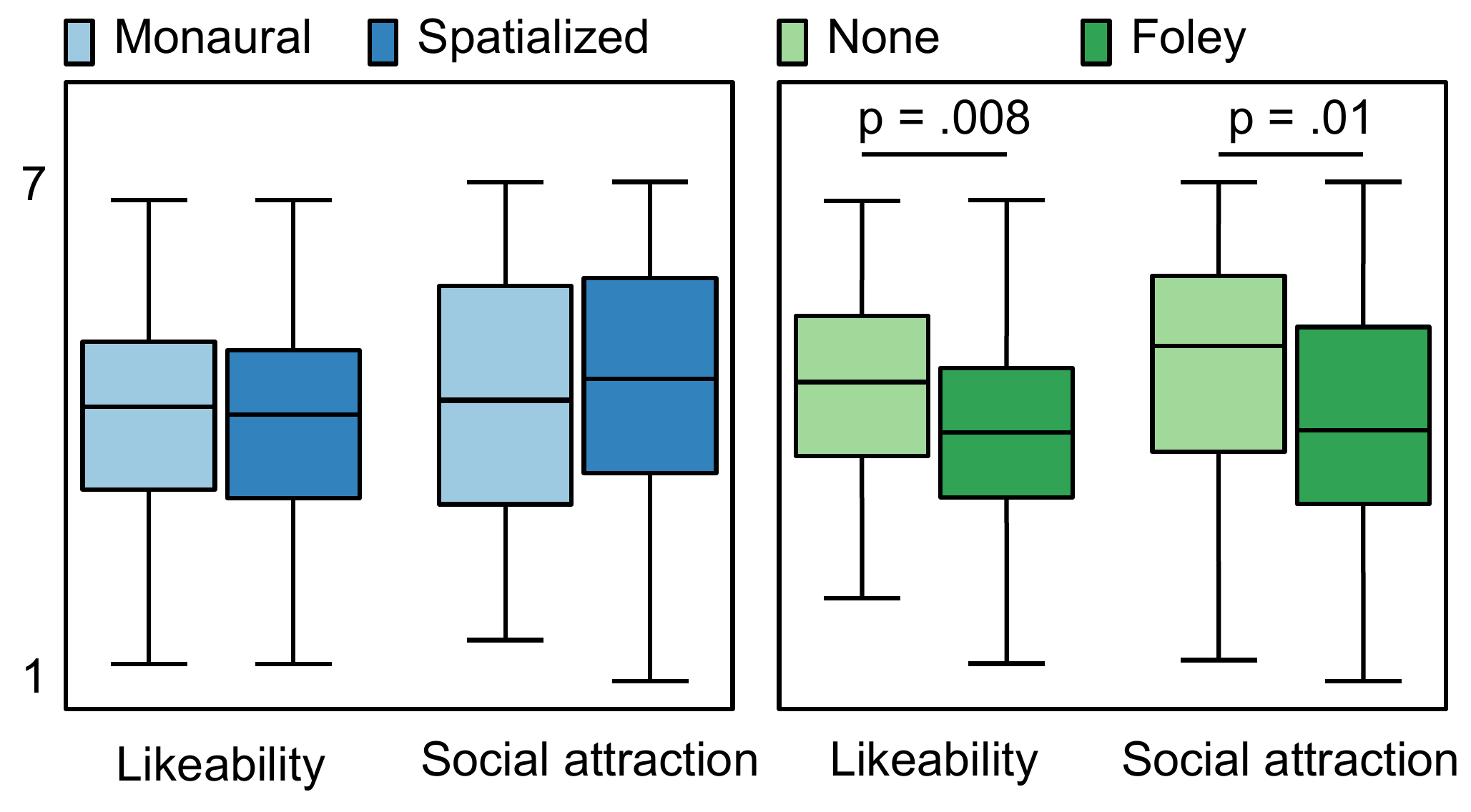}
    \caption{Effect of \textsc{spatialization} and \textsc{Foley} on \emph{likeability} and \emph{social attraction}.}
    \Description{Two boxplot panels showing participant ratings on a scale from 1 to 7 for two categories: Likeability and Social attraction. The left chart compares Monaural with Spatialized conditions. No statistically significant differences are shown. The right chart compares None and Foley conditions. Foley was rated significantly lower for Likeability (p = .008) and Social Attraction (p = .01) compared to the None condition.}
    \label{fig:social-impression}
\end{figure}

\subsection{Social Impression}
We calculated composite scores for \textsl{likeability}~(Cronbach's~\cronbach{0.87}) and \textsl{social attraction}~(Cronbach's~\cronbach{0.89}) from their respective questionnaire items (\autoref{fig:social-impression}).

The ART analysis showed a significant main effect of \textsc{Foley} on both 
\textsl{likeability} (\art{1}{69}{7.45}{=}{0.008}{0.10}) 
and \textsc{social attraction} (\art{1}{69}{6.65}{=}{0.01}{0.09}).
Participants reported lower ratings 
in \textsl{Foley} conditions compared to \textsl{none}, suggesting that \textbf{Foley negatively influenced participants' impressions of the agent}.

While there were no main effects of \textsc{spatialization} on \textsl{likeability} (\art{1}{69}{0.02}{=}{0.9}{.0003}) 
or \textsl{social attraction} (\art{1}{69}{0.31}{=}{0.6}{.005}), 
it interacted significantly with \textsc{Foley} for the latter (\art{1}{69}{5.25}{=}{0.02}{0.07}). Post-hoc tests showed that \textsc{Foley} reduced \textsl{social attraction} in the \textsl{monaural} condition (\pval{=}{0.02}).

\begin{figure}[t]
    \centering
    \includegraphics[width=\linewidth]{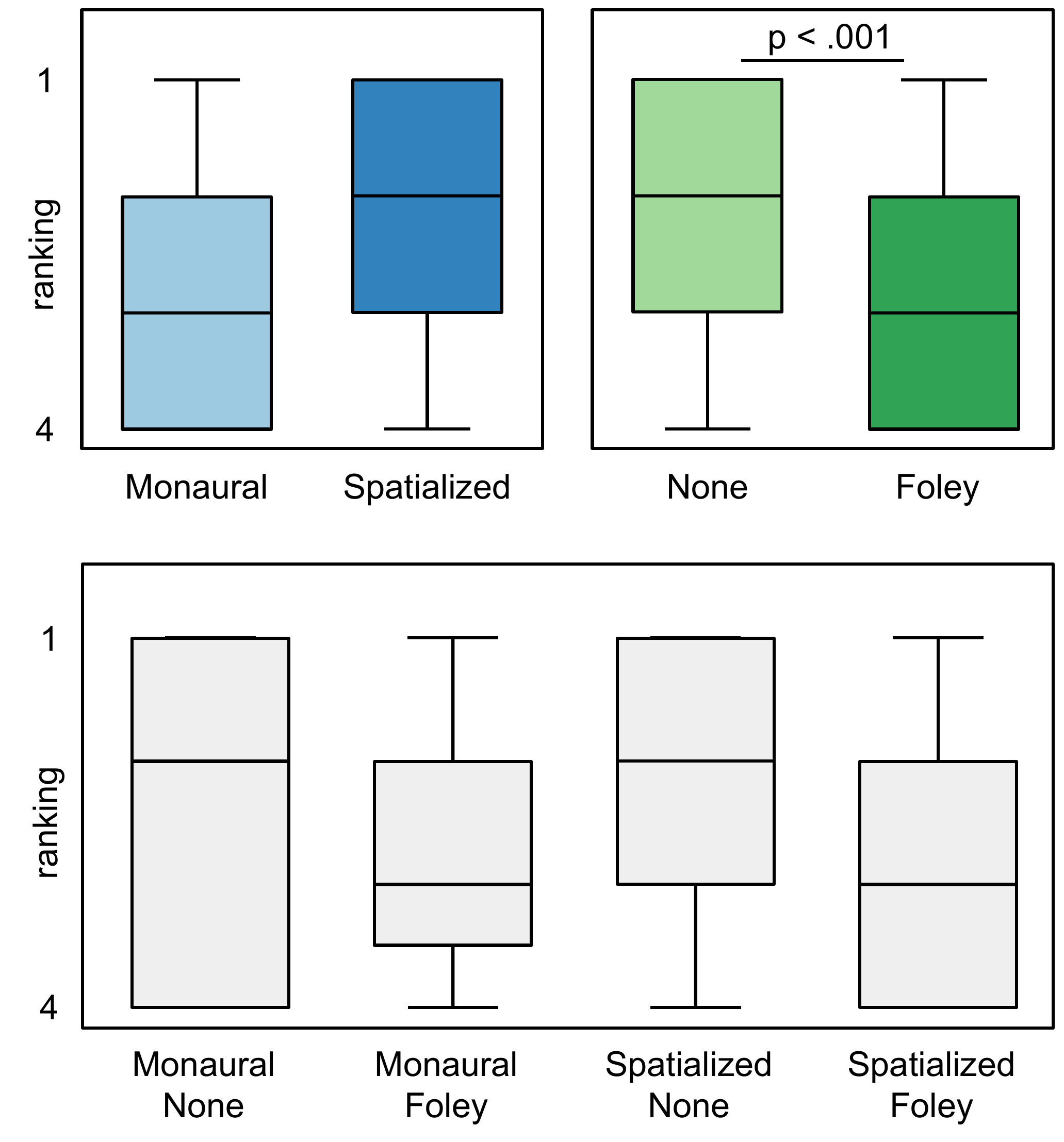}
    \caption{Effect of \textsc{spatialization} and \textsc{Foley} on \emph{preference} rankings (1-Most preferred, 4-Least preferred).}
    \Description{
    Two boxplot panels showing participant preference rankings on a scale from 1 (Most preferred) to 4 (Least preferred).
    The top panel shows main effects. The left chart compares Monaural versus Spatialized. No statistically significant difference is shown. The right chart compares None versus Foley. The None condition was significantly more preferred than the Foley condition.
    The bottom panel shows rankings for all four conditions individually. Spatialized with Foley is ranked highest, followed by spatialized without Foley. Monaural with Foley ranks third, and monaural without Foley ranks lowest. Differences in rankings across the four conditions are not statistically significant.
    }
    \label{fig:preference}
\end{figure}

\subsection{Preference}
The ART analysis showed that participants ranked the conversational agent with \textsl{Foley} below \textsl{none} (\art{1}{69}{6.03}{=}{0.02}{0.08}; Figure~\ref{fig:preference}).
No main effects of \textsc{spatialization} or interactions were observed (all \pval{>}{.05}). 
Overall, these results suggest that \textbf{adding Foley negatively affected perceptions of the agent}.

\subsection{Verbal Behaviors}
\begin{figure}[t]
    \centering
    \includegraphics[width=\linewidth]{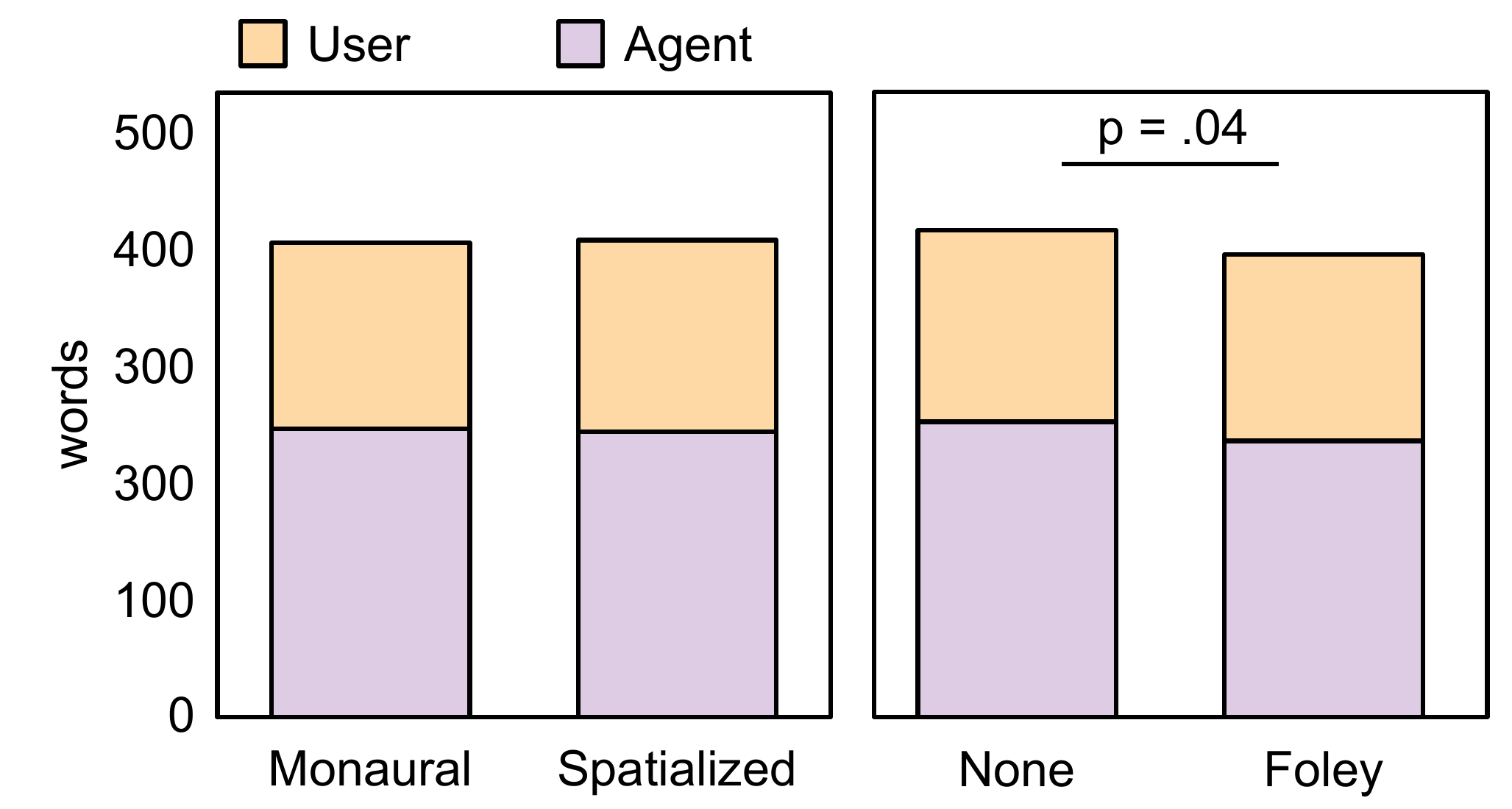}
    \caption{Effect of \textsc{spatialization} and \textsc{Foley} on the number of \emph{words} exchanged in the conversation.}
    \Description{
    Two stacked bar charts showing the average number of words exchanged in the conversation, ranging from 0 to 500 on the y-axis. Each bar is divided into two sections representing the speaker: Agent and User.
    The left panel compares Monaural versus Spatialized conditions. The total height of the bars is nearly identical (approximately 400 words), indicating no statistically significant difference between these conditions.
    The right panel compares the None condition versus the Foley condition. The total number of words exchanged was significantly higher in the None condition compared to the Foley condition (p = .04).
    }
    \label{fig:words}
\end{figure}

On average, conversations comprised 404 words (\sd{52}) exchanged over 21 turns (\sd{5}), including 160 words (\sd{50}) from participants and 244 words (\sd{51}) from the agent (\autoref{fig:words}).

The ART analysis showed that introducing \textsl{Foley} reduced the \textsl{total words} exchanged within the conversation by 20 (\art{1}{69}{4.21}{=}{0.04}{0.06}).
No additional main effects or interaction effects were observed for the other verbal behavior metrics (all \pval{>}{0.05}), including participants’ and the agent’s individual contributions.
These results suggest that the \textbf{introduction of Foley sounds may have slightly dampened conversational engagement}, leading to fewer words being exchanged overall.

\begin{figure}[t]
    \centering
    \includegraphics[width=\linewidth]{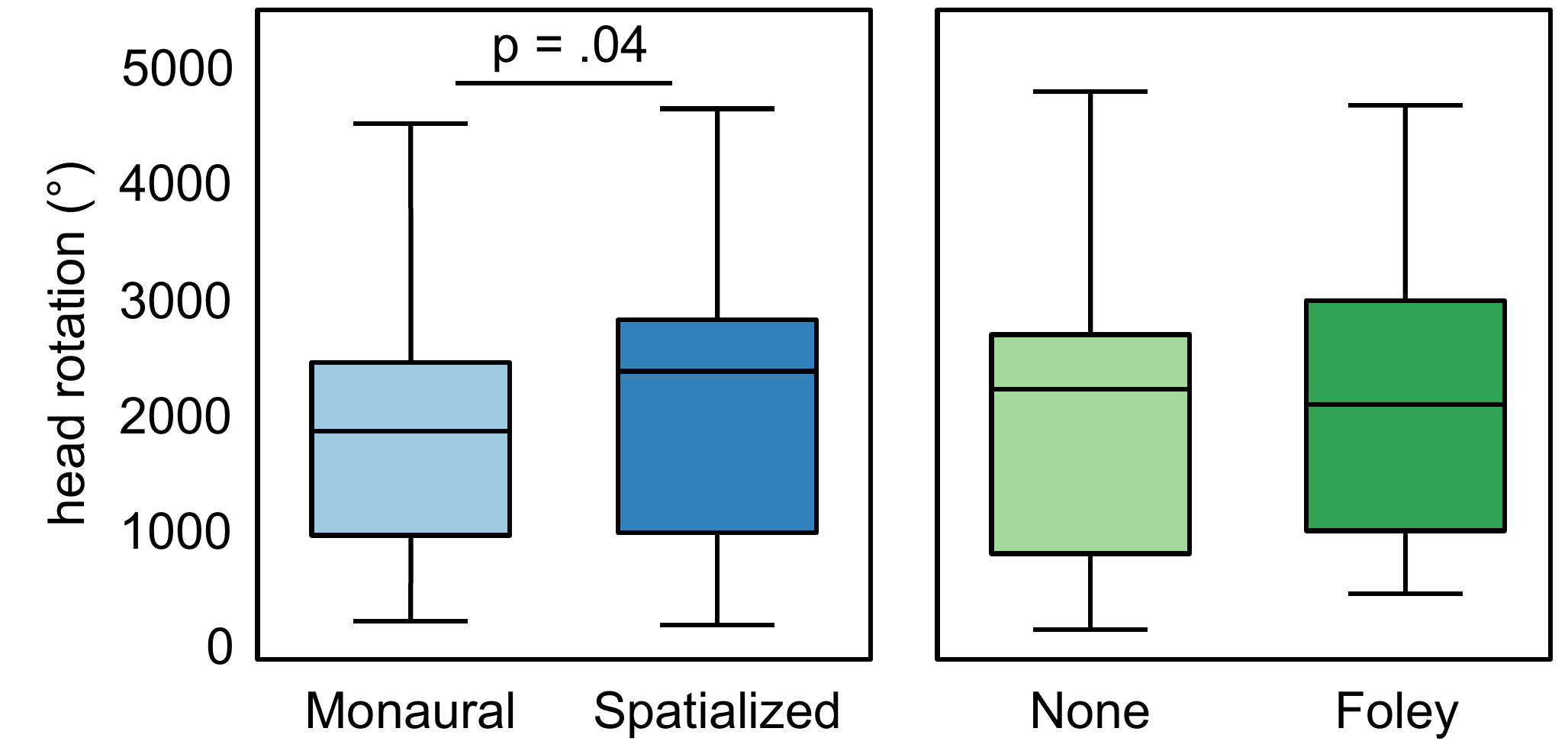}
    \caption{Effect of \textsc{spatialization} and \textsc{Foley} on \textsl{head rotation}.}
    \Description{Two boxplot panels showing the amount of head rotation in degrees across conditions. The panel comparing Monaural versus Spatialized conditions indicates that the Spatialized condition elicited significantly more head rotation compared to the Monaural condition (p = .04). The panel comparing None versus Foley conditions shows a smaller difference that was not statistically significant.}
    \label{fig:head-rotation}
\end{figure}

\subsection{Nonverbal Behaviors}
The ART analysis showed that participants rotated their heads $12\%$ more in the \textsl{spatialized} condition than in the \textsl{monaural} condition~(\art{1}{69}{4.60}{=}{0.04}{0.03}; \autoref{fig:head-rotation}).
No additional main effects or interaction effects were observed for the other nonverbal behavior metrics (all \pval{>}{0.05}). 
These results indicate that \textbf{spatializing the agent’s audio may have led participants to move their heads more frequently during the conversation}.

%% file: sections/qualitative.tex
\subsection{Qualitative Findings}
We analyzed the interview data to gain further insight into participants’ experiences of conversing with an agent under varying levels of \emph{spatialization} and \emph{Foley}.
All interviews were audio-recorded and transcribed. 
We conducted a bottom-up thematic analysis of the transcripts. 
One researcher manually generated 166 open codes, which were iteratively organized into themes. 
These themes were then refined in discussion with other members of the research team. 

\subsubsection{Effect of Spatialization}
13 participants reported that spatialization enhanced their perception of the agent’s presence in the room.
For example, P3 explained that when the audio shifted ``from one ear to another,'' it felt as though the agent was moving ``around the room,'' which made the interaction feel ``more realistic.''
In some cases, the effect was compelling enough to elicit physical reactions: 4 participants described how it prompted them to ``turn to look'' (P8) toward the agent’s perceived location in the room.

However, the perceived strength of the spatialization effects varied among participants.
Although we used a state-of-the-art spatial audio renderer, 
4 participants noted that the spatialization ``wasn't distinct'' (P19), which limited their ability to perceive the agent as moving around the room. 
7 participants reported difficulties identifying the precise location of the agent in the spatialized conditions. 
One such difficulty was distinguishing whether the agent was in the ``background or foreground'' (P9). 
3 participants attributed their confusion to conflicts with what they could see: ``if I'm looking in front of me, I know nothing else is in the room'' (P2). 
These comments reflect the effects of well-documented perceptual limits, despite the technical fidelity of our system, including angular discrimination errors, front–back confusion~\cite{cho2024auptimize}, and distortions due to visual dominance~\cite{lotto2011psychology}.

Participants also highlighted that the agent’s movement behaviors influenced their perception. 
2 participants reported appreciating the movements, 
while 5 others described them as disorienting or distracting. 
As a result of the agent moving around the room, P24 noted that they would occasionally 
``lose track of it.''
To mitigate these issues, P6 suggested anchoring the agent to a single location, such as ``in front of [the user] or a side on the couch sitting area.''

\subsubsection{Effect of Foley}
Responses to the addition of Foley were more polarized. 
10 participants felt that the added sounds contributed to a feeling of co-presence.
For instance, P13 noted that the added sounds ``helped [the agent] to have a connection with what [they] do in the room.''
Especially coupled with spatialization, 
participants felt the effects helped ground the agent within their environment.  
According to P12, spatializing the Foley created a ``good connection'' that tied everything together.

However, while many participants associated the additional sounds with the agent’s interactions in the environment, this was often perceived negatively. 
13 participants felt that the agent seemed distracted:
``it felt like I was in the same room as the agent, but it also felt like the agent was distracted from what I was doing and trying to multitask'' (P20). 
In some cases, these impressions even evoked ``anxiety'' or a sense of ``disrespect'' (P15). 
3 participants perceived the Foley as entirely ``disconnected from the environment'' (P9), 
while 2 others associated them with background noises on a voice call rather than sounds originating from their own room.
As P5 remarked, ``it sounded more like [the agent] was doing something at their place and I just called them.''

Several participants suggested that the added Foley could serve as an effective mechanism for co-presence if used more sparingly: 
``audio cues are helpful, I think they are just too much'' (P1).
P15 suggested that the sounds could be introduced more intelligently by accounting for the conversational context,
 such as during quieter moments rather than in the middle of active dialogue.

\subsubsection{Consideration of Human Conversational Norms}
The more negative impressions invoked by agent movements and Foley can, in part, be attributed to a misalignment with human conversational norms. 
5 participants found the agent's behaviors, moving and interacting with the environment while conversing, implausible in human-to-human interaction. 
As P1 remarked, the agent’s actions, such as pouring water or moving around, are ``not what people do ... when you talk to someone.'' 
P23 shared this sentiment, noting, ``if I’m having a conversation with someone, then I expect them not to be messing with Lego blocks or writing on the board.''

Yet, 2 participants felt that these cues contributed to perceptions of the agent’s humanity.
With spatialization and Foley, P20 described the interaction as feeling more like ``talking to a friend or someone that [they're] in the same space,'' compared to a ``robotic ... customer care'' call when neither cue was available.
Similarly, P3 reasoned that ``people can be doing something in your kitchen or room and still listen to you,'' suggesting that agents, too, should be capable of such peripheral interactions.

%% file: sections/discussion.tex
\section{Discussion}
\label{sec:discussion}
In this paper, we explored how auditory embodiment influences user perceptions of a conversational agent. 
We operationalized auditory embodiment through two manipulations: \textsc{spatializing} the agent’s audio and adding \textsc{Foley}.
In a study,  we investigated how varying these dimensions shaped participants’ experiences.
We will now discuss our findings, design implications, applications, study limitations, and future work.

\subsection{Increasing co-presence}
Our results showed that for a casual conversation scenario, 
spatializing the conversational agent’s audio and adding \textsc{Foley} cues enhanced participants’ perceptions of the agent's co-presence.
These findings are consistent with prior work on the effects of audio on the social presence and perceptions of agents~\cite{immohr2023multimodalspatialaudio,oh2018socialpresence,cummings2022socialpresence}, which has shown that richer auditory representations can positively influence user experiences when interacting with a visually represented avatar or agent. 
Our results also align with the conceptualization of presence as ``realism''~\cite{lombard1997heartofitall}, which posits that presence concerns the degree to which a medium can produce accurate representations of events, objects, and people. 
Building on this work, our study empirically demonstrates that auditory cues can, independent of any visual representation, convey a sense of co-presence with a conversational agent situated within the user’s physical space.

\subsection{Co-present but distracted}
While \textsl{spatialization} and \textsl{Foley} enhanced participants’ perception of co-presence, they came with a cost. 
Subjective ratings suggest that the addition of \textsl{Foley} contributed to perceptions of the agent as less attentive to the conversation and reduced its perceived message understanding, likeability, and social attraction. 
It also decreased the total number of words exchanged between the participant and the agent, indicating lower conversational engagement.
Sometimes, because the agent was scripted to move around the room, \textsl{spatialization} also contributed to participants’ negative impressions of the agent, as reflected in the interview comments.

Our findings suggest that these effects may stem from the way \textsl{spatialization} and \textsl{Foley}, by introducing a bodily presence, inadvertently encouraged participants to anthropomorphize the system.
As a computational system, our agent cannot technically be ``distracted'' or ``not pay attention,'' but our participants projected human attentional qualities and conversational norms onto it.
This largely aligns with the computers as social actors paradigm~\cite{nass1994socialactors} and the Media Equation theory~\cite{reeves1996media}. 
Hence, participants' more negative perceptions of the agent as a result of \textsl{spatialization} and \textsl{Foley} can, in part, be explained by a misalignment between the agent's embodied behaviors and social expectations. 
Just as people would be perceived as distracted and disrespectful if they multitask during a conversation, participants likewise interpreted the agent’s concurrent movements and activities as signs of inattention.

\subsection{Perceptual limitations}
Our results highlighted how known perceptual limitations constrained participants’ experiences of the agent’s embodiment.
First, while sounds offer rich spatial cues, humans are prone to several auditory ambiguities. Prior work has shown that within controlled environments, localization error for loudspeakers can reach up to $\pm10^{\circ}$\cite{blauert1997spatial}. People are also notably poor at distinguishing between sound sources located in front of versus behind the head~\cite{cho2024auptimize}. Qualitative feedback from the interviews suggests that both of these perceptual limitations manifested in participants’ experiences and, to some extent, hindered their ability to perceive the agent as fully co-present within their room.
Second, multisensory perception literature suggests that people are generally visually dominant: when audio and visual cues conflict, the visual system tends to override the auditory~\cite{lotto2011psychology}. Our results similarly showed the difficulty of convincingly representing an agent when it is visually absent.
We believe that the addition of \textsl{Foley}, and the resulting increase in co-presence, is one means to mitigate poor localization performance.

\subsection{Design Considerations}
Overall, our results suggest that \textsl{spatialization} and \textsl{Foley} can be effective mechanisms for enhancing users’ sense of co-presence with conversational agents; however, they also introduce trade-offs, particularly in perceptions of agent attention, message understanding, and social impressions.
Based on our findings, we discuss several considerations for the design of auditory agents. 

\subsubsection{Alignment with social norms}
When a conversational agent is auditorily embodied, users expect it to exhibit human-like behaviors in its movements and interactions with the environment. These behaviors are evaluated against familiar social norms, and violations of those norms will be perceived negatively.
For instance, just as another person would be considered rude if they wandered around the room and arbitrarily interacted with objects during a conversation, an agent whose audio cues signal similar behavior may be perceived as inattentive or disrespectful.
Consequently, auditory embodiments, including choices about spatialization and Foley sounds that convey the agent’s incidental interactions, must be not only physically plausible within the user’s environment, but also sensitive to social norms.

\subsubsection{Accounting for perceptual limitations}
The effectiveness of spatialization and Foley as mechanisms for conveying auditory embodiment is contingent on perceptual constraints. 
For spatialization, designers must consider the limitations of auditory localization, including angular discrimination errors and front–back confusion. 
These limitations make it difficult for users to perceive subtle or rapid shifts in the agent’s position. 
To better support users in situating the agent within their environment, designers may opt for slower and more pronounced movements.
For Foley, visual dominance can overshadow incidental auditory cues. 
When Foley is played for an object the user is directly looking at, our results suggest that users will likely struggle to reconcile the auditory cue with the absence of a visible agent. 
Foley may be more effectively leveraged to situate the agent in locations outside the user’s direct field of view, where the effects of visual dominance are reduced.

\subsection{Applications}
Our results suggest that spatialization and Foley positively influence co-presence, but reduce attention and other social factors. 
We believe these implications extend to a wide range of applications involving conversational agents, including the auditory design of social AI companions~\cite{fang2025aihumanbehaviorsshape}. 
In these contexts, spatialized vocal presence and incidental Foley cues may offer a means of enhancing an agent’s social presence, which in turn can shape users’ perceptions of its usefulness and sociability~\cite{merrill2022companions}. 
These mechanisms may also support new forms of situated storytelling, enabling richer narrative experiences similar to those explored by \citeauthor{li2022arstorytelling}~\cite{li2022arstorytelling}, where auditory cues help anchor characters and events within the user’s environment.
Beyond benefits tied to social presence, we speculate that auditory embodiment may also offer advantages analogous to visual embodiment for spatial tasks, such as guiding navigation~\cite{bohus2019situated}. 
By providing spatially grounded directional cues, an auditorily embodied agent may support more efficient wayfinding.

At the same time, our findings highlight important boundaries. 
Embodiment may not always be necessary, particularly for interactions that serve primarily transactional or functional goals~\cite{cheng2025surrogates}, where additional social cues may introduce unnecessary cognitive load. Moreover, embodiment is not universally desirable: prior work has shown that it can foster over-reliance and other problematic social dynamics~\cite{de2016ethical}. Here, our results point to opportunities in intentionally leveraging social cues of inattention or distraction to introduce ``seams''~\cite{chalmers2003seamful} in the agent interaction. 
Such seamful cues can act as gentle reminders of the system’s non-human nature, helping users detach and exercise their own judgment regarding the content of the conversation.

\subsection{Study Limitations}
\label{sec:limitations}
Our study is subject to several limitations, which we discuss below.

\subsubsection{Conversation Task}
First, we only explored a constrained conversational context in our study. Participants were instructed to engage in a casual dialogue on pre-defined topics that were designed to be self-relevant yet not emotionally charged, in order to maintain experimental control over emotional valence and arousal. 
While a valuable first step, how exactly the effects of auditory embodiment translate to other conversational scenarios remains unclear and presents opportunities for further exploration. 
For instance, several participants suggested in interviews that for task-oriented conversations, auditory embodiment may offer less value.
Future work should therefore consider examining the interaction between auditory embodiment and the goals of the conversation, \eg~transactional vs. social~\cite{cheng2025surrogates}.
Similarly, while our results did not show any effect on affective understanding or interdependence, it remains an open question whether these outcomes might change as a function of the intimacy or affective tone of the conversation. 
Finally, extending beyond dyadic conversations, future work could investigate the value of auditory embodiment in group interactions, particularly in hybrid settings~\cite{mirrorblender2021gronbaek}.

\subsubsection{Agent Behavior}
In our study, the agent’s embodied behaviors within the environment were pre-scripted. 
These behaviors were designed with the objectives of maintaining ecological plausibility and providing perceptual grounding, while remaining mostly ambient to the conversation.
Our results suggest that the pre-scripted behaviors we designed increased feelings of co-presence but reduced attention and other social factors. 
A primary reason for this appears to be that the agent’s behaviors were misaligned with the social norms of the conversational context.
This raises the question of how to reconcile this gap. 
Here, it is worth noting that the conversational agent in our study was wholly unaware of its embodied behaviors.
Similarly, its activities within the physical environment were agnostic to whether, when, and what the agent was communicating. 
This disconnect between dialogue and embodiment may have contributed to the perceived misalignment with social norms. 
Future work could therefore explore system architectures in which an agent’s conversational and embodied behaviors are more tightly coupled, enabling the system to coordinate speech, movement, and environmental actions in socially appropriate ways.

Our interview results provide some early insights into what may be considered more desirable behavior. 
Several participants suggested reducing the agent's movements and activities.
However, it remains unclear whether such subtle cues are sufficient to support a stronger sense of co-presence.
In particular, reducing an agent’s movement may diminish the spatial information needed for localization, and reducing its activities may similarly limit opportunities to situate the agent within the user’s environment.
Future work could therefore examine how different degrees of movement, from static anchoring to more dynamic environmental interaction, shape user perceptions of presence, attentiveness, and social appropriateness.

\subsubsection{Beyond Focused Conversations}
In our study, participants were seated and asked to converse without any competing demands. 
However, just as the agent in our study was scripted to perform activities in parallel, users themselves may also be engaged in other tasks in real life.
This raises questions about how the effects of auditory embodiment may interact with additional task requirements. 
On the one hand, additional cognitive load may reduce the user’s audio perception and localization abilities.
On the other hand, the user's task may mediate the social acceptability of the agent's multitasking. 
Future work could consider varying whether the agent is completing independent tasks or engaged in the same tasks as the user.
One of the authors occasionally wishes for company while doing chores.
In this scenario, Foley cues indicating that the agent is sharing the burden may be welcomed.

\subsubsection{Sample size and generalizability} 
For our study, we recruited 24 participants from a university context. Although we believe this sample size was sufficient for an initial investigation, replicating the study with a larger and more diverse participant pool will be important for strengthening the generalizability of our findings.

In addition, the study was conducted in a controlled laboratory environment. From a perceptual perspective, people’s ability to associate sounds with locations partly depends on their cognitive map, which develops with familiarity over time. While we included a familiarization step, participants’ associations with sounds may have been stronger in more personally familiar or ecologically valid contexts. 
More broadly, it remains an open question how auditory embodiments will function in diverse real-world settings, which are often dynamic and filled with competing sound sources.

\subsection{Towards Deployment}
Although our main contribution is an empirical investigation of auditory embodiment, we see value in extending our experimental implementation into a deployable system. 
Auditory embodiment ultimately relies on three technical capabilities: 
3D spatial capture of users and their environment,
spatial audio rendering, 
and environment-audio retrieval or synthesis. 
In our implementation, we used an OptiTrack system for user tracking, 
the Meta XR Audio SDK for spatial audio rendering, 
and a curated library of pre-recorded environmental audio to approximate in-situ ambient sound. 
While effective for controlled studies, 
our tracking apparatus and reliance on pre-recorded audio sequences pose clear challenges for scaling the system beyond a single environment. 
We believe these components can be replaced with more portable and adaptive alternatives. 
For instance, many modern head-mounted displays, such as the Meta Quest, already provide on-device user tracking and basic environment understanding~\cite{metaxrsceneunderstanding,metaxrtracking}. 
This information can support the automatic synthesis and spatial placement of contextually aligned Foley sounds~\cite{schutz2025sonifyanything,su2024sonifyar,rowles2025foleycontrol,lin2023soundify}.
Future work should integrate these capabilities into a mobile end-to-end system for in-the-wild deployment, enabling context-aware auditory embodiment in everyday environments.

%% file: sections/conclusion.tex
\section{Conclusion}
In this work, we explore how auditorily embodying a conversational agent through spatializing its audio and introducing incidental Foley audio affects social presence and perception. 
Our results from an experiment with 24 participants indicate that while both spatialization and Foley enhance feelings of co-presence, Foley reduces perceived attention, message understanding, likeability, and social attraction.
As conversational agents become increasingly pervasive, our work highlights auditory embodiment as a promising approach for enabling richer interactions, while underscoring the need to carefully consider the trade-offs it may introduce.

%% file: sections/acknowledgements.tex
\begin{acks}
We thank all involved peers, participants, and anonymous reviewers, especially Shwetha Rajaram, Portia Wang, Yujie Tao, and Shannon Yeung for their input throughout the project. 
Yi Fei Cheng was supported by the \href{https://croucher.org.hk/}{Croucher Foundation}.
Dr. Andrea Bianchi was supported by the National Research Foundation of Korea (NRF) grant funded by the Korea government (MSIT) (RS-2024-00337803).
Dr. Anusha Withana is a recipient of an Australian Research Council Discovery Early Career Award (DECRA) - DE200100479, funded by the Australian Government.
\end{acks}

%% file: sections/appendix.tex
\appendix
\section{Conversation Topics}
\label{appendix:topics}
For our conversation task, participants randomly selected from the following topics, adapted from~\citeauthor{fang2025aihumanbehaviorsshape}~\cite{fang2025aihumanbehaviorsshape}:
\begin{itemize}
    \item Let's chat about the best gift I ever received.
    \item Let's chat about a concert or show I went to that was memorable.
    \item Let's chat about my favourite holiday.
    \item Let's chat about the best show I've watched in the past few months.
    \item Let's chat about what a perfect day would look like for me.
    \item Let's chat about how I celebrated a recent holiday.
    \item Let's chat about the best book I've read in the past year. 
\end{itemize}

\section{Voice \rrh{Configuration}}
\label{appendix:voice-config}
The audio model used for our conversation task was prompted with the following system instruction: 
\begin{verbatim}
You are an AI companion that a user is going to engage 
in a casual conversation with. 
\end{verbatim}

\section{Post-condition Questionnaire Items}
\label{appendix:questionnaire-items}
Our post-condition questionnaire for the participant included a subset of the Networked Minds Social Presence Inventory (NMSPI)~\cite{biocca2003networked} evaluating co-presence (C), attentional allocation (AA), message understanding (MU), affective understanding (AU), and affective interdependence (AI). 
In addition, we collected likeability (L) and social attraction (SA) judgements, adapting questions from \citeauthor{fauville2022impression}~\cite{fauville2022impression} and \citeauthor{tao2025audiopersonas}~\cite{tao2025audiopersonas}. All items used 7-point Likert scales (1-Strongly disagree, 7-Strongly agree). 

\begin{itemize}
    \item[C1] I often felt as if the agent and I were in the same room together. 
    \item[C2] I think the agent often felt as if we were in the same room together.
    \item[C3] I was often aware of the agent in the room.
    \item[C4] The agent was often aware of me in the room. 
    \item[C5] I hardly noticed the agent in the room. 
    \item[C6] The agent didn't notice me in the room.
    \item[C7] I often felt as if we were in different places rather than the same room. 
    \item[C8] I think the agent often felt as if we were in different places rather than together in the same room.  
    \item[AA1] I was easily distracted from the agent when other things were going on. 
    \item[AA2] The agent was easily distracted from me when other things were going on. 
    \item[AA3] I remained focused on the agent throughout our interaction.
    \item[AA4] The agent remained focused on me throughout our interaction. 
    \item[AA5] The agent did not receive my full attention.
    \item[AA6] I did not receive the agent's full attention.
    \item[MU1] My thoughts were clear to the agent. 
    \item[MU2] The agent's thoughts were clear to me. 
    \item[MU3] It was easy to understand the agent. 
    \item[MU4] The agent found it easy to understand me.
    \item[MU5] Understanding the agent was difficult. 
    \item[MU6] The agent had difficulty understanding me. 
    \item[AU1] I could tell how the agent felt.
    \item[AU2] The agent could tell how I felt. 
    \item[AU3] The agent's emotions were not clear to me. 
    \item[AU4] My emotions were not clear to the agent.
    \item[AU5] I could describe the agent's feelings accurately.
    \item[AU6] The agent could describe my feelings accurately. 
    \item[AI1] I was sometimes influenced by the agent's moods.
    \item[AI2] The agent was sometimes influenced by my moods.
    \item[AI3] The agent's feelings influenced the mood of our interaction. 
    \item[AI4] My feelings influenced the mood of our interactions. 
    \item[AI5] The agent's attitudes influenced how I felt. 
    \item[AI6] My attitudes influenced how the agent felt. 
    \item[L1] I think this agent is attractive.
    \item[L2] I think this agent is competent.
    \item[L3] I think this agent is extroverted.
    \item[L4] I think this agent is likeable.
    \item[L5] I think this agent is trustworthy.
    \item[SA1] I like this agent.
    \item[SA2] I get along with this agent.
    \item[SA3] I would enjoy a casual conversation with this agent again.
    \item[SA4] I think this agent is friendly. 
\end{itemize}